\documentclass[12pt]{iopart}
\usepackage[dvips]{graphicx}
\usepackage[dvips]{color}
\usepackage{iopams}
\usepackage{mhequ}

\def\m#1{\mathbf{#1}}

\def\g{{\mathcal{G}}}

\def\e{\varepsilon}
\def\kB{k_\mathrm{B}}
\def\rhs{{\em r.h.s.} }
\def\wrt{{\em w.r.t.} }

\def\an{\alpha_n}

\def\mS{\mathbf{S}}

\def\mU{\mathbf{U}}
\def\mW{{W}}
\def\mC{\mathbf{c}}
\def\mV{\mathbf{v}}
\def\mY{\mathbf{y}}
\def\mZ{\mathbf{z}}

\def\mS{\mathbf{s}}
\def\nZ{{\mathcal{Z}}}

\def\nV{{\mathcal{V}}}
\def\nPsi{{\Psi}}
\def\nZ{{\mathcal{Z}}}
\def\nS{\mathcal{S}}
\def\l{\lambda}
\def\g{\gamma}
\def\w{\omega}
\def\z{\zeta}
\def\ddelta{\tilde{\delta}}

\def\Lop{\mathcal{L}}
\def\KV{{\mathcal{K}}_{\mathcal{V}}}

\newcommand{\cor}[1]{\langle #1\rangle}

\newcommand{\pderiv}[1]{\frac{\partial #1}{\partial t}}

\begin{document}

\title[Dynamics of anomalous heat transport]
{Nonequilibrium dynamics of a stochastic model of anomalous heat transport}
\author{Stefano   Lepri,  Carlos   Mej\'ia-Monasterio\footnote{Present 
    address: University of Helsinki, Department of Mathematics and Statistics
    P.O. Box 68 FIN-00014, Helsinki, Finland}
  and  Antonio   Politi} 
\address{Istituto dei  Sistemi Complessi, Consiglio  Nazionale delle Ricerche,
via Madonna del piano 10, I-50019 
  Sesto     Fiorentino,    Italy}    \eads{\mailto{stefano.lepri@isc.cnr.it}, 
\mailto{carlos.mejia@helsinki.fi}, \mailto{antonio.politi@isc.cnr.it}}
\begin{abstract}
  We study the dynamics of covariances in a chain of harmonic
  oscillators with conservative noise in contact with two stochastic
  Langevin heat baths. The noise amounts to
  random collisions between nearest-neighbour oscillators that
  exchange their momenta.  In a recent paper, [S Lepri et al.
  \textit{J. Phys. A: Math. Theor.} 42 (2009) 025001], we have studied
  the stationary state of this system with fixed boundary conditions,
  finding analytical exact expressions for the temperature profile and
  the heat current in the thermodynamic (continuum) limit.  In this
  paper we extend the analysis to the evolution of the covariance
  matrix and to generic boundary conditions. Our main purpose is to
  construct a hydrodynamic description of the relaxation to the
  stationary state, starting from the exact equations governing the
  evolution of the correlation matrix. We identify and adiabatically
  eliminate the fast variables, arriving at a continuity equation for
  the temperature profile $T(y,t)$, complemented by an ordinary
  equation that accounts for the evolution in the bulk. Altogether, we
  find that the evolution of $T(y,t)$ is the result of fractional
  diffusion.
\end{abstract}
\submitto{\JPA}

%%%%%%%%%%%%%%%%%%%%%%%%%%%%%%%%%%%%%%%%%%%%%%%%%%%%%%%%%%%%%%%%%%%%%%%%%%%%%%%
%%%%%%%%%%%%%%%%%%%%%%%%%%%%%%%%%%%%%%%%%%%%%%%%%%%%%%%%%%%%%%%%%%%%%%%%%%%%%%%
\section{Introduction.}
\label{sec:intro}

The chain of harmonic oscillators maintained out of equilibrium by
means of stochastic heat reservoirs, is one of the few systems for
which the nonequilibrium invariant state has been obtained rigorously
\cite{RLL67}.  However, due to the lack of any phonon scattering
mechanism, its heat conductivity $\kappa$ diverges linearly with the
size of the chain. Consequently, the Fourier's law of heat conduction
$J_Q=-\kappa \nabla T$, relating the heat flux $J_Q$ with the imposed
temperature gradient $\nabla T$ does not hold.  As a matter of fact,
the integrability of the harmonic chain makes impossible for the
system to support a temperature gradient.  It was already recognized
by Debye that the presence of nonlinearities in the dynamics is a
necessary condition for the occurrence of a {\it normal transport},
i.e. a finite heat conductivity in the thermodynamic limit. However,
years later Fermi, Pasta and Ulam would found that nonlinear dynamics
does not necessarily induces a statistical behaviour \cite{FPU}.

In the last decade, numerical simulations and analytic arguments have
contributed to clarify the role of nonlinearities in the thermodynamic limit
\cite{LLP03,EPRB99,BLL04,LS06,BK07}.  In some studies, anharmonicities
have been introduced by means of self-consistent local thermostats,
\cite{BLL04,BRV70,DR06}.  Further attempts to derive Fourier's law in
deterministic systems have been reported (see review papers
\cite{BLR00,DHAR09} and references therein).  However, no rigorous
derivation exists of the necessary and sufficient conditions for the
validity of Fourier's law.  Moreover, there is still a number of open
questions concerning the steady state, such as the role of Boundary
Conditions (BC in the following), while the convergence towards the
stationary state is even less explored.

Stochastic models are rather useful in that they can effectively
reproduce the evolution of deterministic nonlinear systems, while
allowing for analytic solutions. Bolsterli, Rich and
Visscher considered harmonic chains in which each oscillator is in
contact with a stochastic thermal reservoir \cite{BRV70}. Then, the
stationary state is obtained assuming a self-consistent condition,
namely the energy current between the local reservoirs and the
respective oscillator is zero. Recently, it has been proved by Bonetto
et. al. that this linear model leads to a Gaussian invariant measure
and the temperature profiles are linear \cite{BLL04}. A drawback of
this model is that, strictly speaking, energy is not conserved by the
bulk dynamics (see also \cite{Davies78} where energy current from the
reservoirs becomes zero in the long time limit, and \cite{DR06} for
their treatment in terms of nonequilibrium Green function formalism).
Another model that can be explicitly solved is the
Kipnis-Marchioro-Presutti (KMP) lattice model, in which stochastic
collisions mix the energy of neighboring particles, conserving the
total energy, but not the momentum \cite{KMP82}.  This model satisfies
Fourier's law and a linear temperature profile is obtained. Energy
conserving stochastic noise has been also used in lattice model
systems, as natural generalizations of KMP and of the single exclusion
process (SEP) \cite{GKR07}.

In the same spirit of KMP, Basile et al. have recently
studied a harmonic chain in the presence of random collisions among triples
of nearest-neighbour oscillators \cite{BBO06}. This latter process, which
conserves both energy and momentum, amounts to a diffusion on the energy shell.
In this system, it is proven that the energy-current autocorrelation decays
as $t^{-1/2}$ and thereby heat conductivity diverges with the size of the
system $N$, as $\kappa\sim N^{1/2}$ \cite{BBO06}.
More recently, Jara et al. have
studied the relationship between anomalous heat transport and
fractional diffusive processes \cite{JKO08}.  They find that in the
infinite system, the dynamics leading to anomalous transport is
obtained from a Levy stable process that corresponds to a space-time
scaling given by the fractional diffusive operator
$\partial_t-\nabla^{3/2}$, where $\nabla$ is the gradient operator.

In a recent paper \cite{LMMP09}, we have studied a harmonic chain with
both energy- and momentum-conserving  noise (and fixed BC). The model,
a slight variant of the one introduced in \cite{BBO06}, is amenable to
analytical calculations,  as  the evolution  equations for  the
covariance  matrix  are linear  (see  also  \cite{DLLP08}).  Taking  a
suitable continuum  limit, recalled in  section \ref{sec:change}, we
obtained a solution for the  covariance matrix in the stationary state
$t\rightarrow\infty$. From  this solution we  derived exact
expressions for the temperature  profile and the heat current, finding
that the  heat conductivity  diverges as in  \cite{BBO06}. Remarkably,
the   temperature   profile  (equation   (8)   in  \cite{LMMP09})   is
parameterfree,  suggesting that  it  may represent  a  wider class  of
systems. To  our knowledge, this is  the first example  of an analytic
expression for  the temperature profile  in a system  characterized by
anomalous heat  transport and confirms that a nonlinear shape persists
in the thermodynamic limit.

In  this  and in  the  companion  paper  \cite{partII} we  extend  the
previous analysis to the nonstatonary case, obtaining a hydrodynamic
equation ruling the relaxation of covariance matrix towards the stationary
state. More precisely, here we analytically investigate the continuum limit,
by progressively eliminating the fast variables. The second part contains a
detailed numerical analysis of all aspects that are not (easily) accessible to
an analytic investigation, including the case of free BC, and the
estimate of finite-size effects. 
At variance with normal heat conduction, the role of BCs is very important
in the anomalous case. For instance, in disordered chains  of linear
oscillators,  the  same  system   may  even   behave  as   a  thermal
superconductor or  as an insulator,  by simply switching from  free to
fixed BC \cite{LLP03}. In the present context, we  find that in the
thermodynamic limit and given bath temperatures, fixed and free BC give rise to
the same scaling behaviour, but macroscopically different values of the heat
flux. Even more surprising, we find that the stationary value
of the  heat flux varies the coupling strength with the heat baths only for
free BC.  

The numerical analysis \cite{partII} demonstrates that these effects are caused
by boundary layers, where the scaling behaviour of some observables changes
with respect to the bulk. This phenomenon, which is associated with strong
deviations from local equilibrium, hinders the development of an analytical
solution in the general case. Nevertheless, for fixed BC, we demonstrate that 
the boundary layers do not affect the relevant physical observables, and an
explicit solution can be obtained for the evolution of the temperature profile.
Our results, obtained for large but finite chains, are consistent with those
derived directly in the infinite-size limit \cite{JKO08}. This was not a priori
granted in the presence of long-range correlations.

This paper is organized as follows. In Section \ref{sec:model}, we
define the stochastic model and introduce the main notations. More
precisely, in Section \ref{sec:cov}, we introduce the covariance
matrix, by adopting a different definition with respect to
\cite{LMMP09}, so as to be able to treat both free and fixed BC. The
corresponding ordinary differential equations are derived in sections
\ref{sec:cov} and \ref{sec:boundaryeqs}, with reference to the bulk
and boundaries, respectively. In Section~\ref{sec:change} we perform a
further change of variables to simplify the treatment of the continuum
limit, discussed in section \ref{sec:expansion}. There, we define the
smallness parameter, $\varepsilon = 1/\sqrt{N}$ ($N$ is the chain
length) and thereby map the discrete spatial indices $i$, and $j$ of
the correlation matrices onto two continuous variables (continuum
limit) $x$ and $y$. Moreover, we introduce an Ansatz for the scaling
behaviour of the different variables, as suggested by the numerical
analysis presented in the companion paper \cite{partII}. The internal
consistency of the resulting equations confirms {\it a posteriori} the
correctness of our initial choice. In section \ref{sec:results}, we
anticipate the main results to allow the reader appreciating them
without being distracted by the technical details. The derivation of
the partial differential equations and the elimination of the fast
degrees of freedom is
illustrated in sections \ref{sec:dynamics}, \ref{sec:bc}, with
reference to the bulk and boundary dynamics, respectively. The
relaxation towards the stationary state is then discussed in
Section~\ref{sec:relax}. The numerical computation of the spectrum of
the covariance evolution operator corroborates the analytical results.
Some concluding remarks are finally presented in
Section~\ref{sec:concl}.

%%%%%%%%%%%%%%%%%%%%%%%%%%%%%%%%%%%%%%%%%%%%%%%%%%%%%%%%%%%%%%%%%%%%%%%%%%%%%%%
%%%%%%%%%%%%%%%%%%%%%%%%%%%%%%%%%%%%%%%%%%%%%%%%%%%%%%%%%%%%%%%%%%%%%%%%%%%%%%%
\section{Stochastic model.}
\label{sec:model}

We consider a homogeneous chain of $N$ harmonic oscillators of unit
mass and frequency $\omega$. The first and $N$-th oscillators are
coupled to Langevin heat baths at different temperatures. The
equations of motion of this system are
\begin{equation} \label{eq:eqs-motion}
\begin{array}{rcl}
\dot q_n & \ = \ & p_n \\
\dot p_n & \ = \ & \omega^2 (q_{n+1} - 2q_n + q_{n-1}) + 
 \delta_{n,1}(\xi^+ - \lambda \dot q_1) + \delta_{n,N}(\xi^- -\lambda \dot
 q_N)  \ .
\end{array} 
\end{equation}
Here $p_n$ and $q_n$ are the momentum and displacement (from its
equilibrium position), of the $n$-th oscillator and $\xi^\pm$ are
independent Wiener processes with zero mean and variance $2\lambda k_B
T_\pm$, 
where $k_B$ is the Boltzmann constant and $\lambda$ is the
coupling constant.  Additionally, the deterministic dynamics is
perturbed by random binary collisions between nearest-neighbour
oscillators occurring at a rate $\gamma$.  These collisions are
defined so that total momentum and energy are conserved. This type of
stochastic noise is known in the literature as {\it conservative
  noise}.

The phase-space probability density $P(\vec{q},\vec{p},t)$ evolves
according to the equation
\begin{equation} \label{eq:F-P}
\pderiv{P} \ = \ \left(\Lop_0 + \Lop_{coll}\right) P \ ,
\end{equation}
The first term corresponds to the usual Liouville generator of the
dynamics, acting on the probability density as
\begin{equation} \label{eq:L0}
\Lop_0 P = \sum_{i,j} \left(\m{a}_{ij} \frac{\partial x_j P}{\partial x_i} +
\frac{\m{d}_{ij}}{2} \frac{\partial^2 P}{\partial x_i \partial x_j}\right) \ , 
\end{equation}
with the $2N$ vector $\m{x} = (q_1, q_2, \ldots, q_N, p_1, p_2, \ldots, p_N)$,
and the $2N \times 2N$ matrices $\m{a}$ and $\m{d}$ are
\begin{equation} \label{eq:mat-0}
\m{a} = \left(
\begin{array}{cc}
\m{0} & -\m{1} \\
\omega^2\m{g} & \lambda\m{r}
\end{array}
\right) \  ; \ \quad 
\m{d} = \left(
\begin{array}{cc}
\m{0} & \m{0} \\
\m{0} & 2\lambda k_B T(\m{r} + \eta\m{s})
\end{array}
\right) \,
\end{equation}
where $T$ is the average temperature $(T_++T_-)/2$, $\eta$ is the
relative temperature difference ($\eta= (T_+-T)/T$), $\m{0}$ and
$\m{1}$ are the null and unit $N \times N$ matrices,
\begin{equation} \label{eq:mat-1}
\m{r}_{ij} \ = \ \delta_{i,j}\left(\delta_{i,1} + \delta_{i,N}\right) \ ,
\quad \ \m{s}_{ij} \ = \ \delta_{i,j}\left(\delta_{i,1} - \delta_{i,N}\right)
\ , 
\end{equation}
and $\m{g}$ is the negative of the discrete Laplacian
\begin{equation} \label{eq:mat-2}
\m{g}_{ij} \ = \ 2\delta_{i,j} - \delta_{i+1,j} - \delta_{i,j+1} \ .
\end{equation}
The stochastic collision generator is
\begin{equation} \label{eq:Lcoll}
\Lop_{coll} P = \gamma \sum_{j=1}^{N-1} \left[
P(\ldots ,p_{j+1},p_j, \ldots) - P(\ldots ,p_j, p_{j+1}, \ldots) \ \right] \ .
\end{equation}
Each term  in the  sum expresses the  probability balance for  each elementary
process  in which the momenta of each pair $j,j+1$ are exchanged with a rate
$\gamma$.

In this  paper we consider  either free or  fixed boundary conditions  and the
choice adopted will be explicitly stated wherever needed.

%%%%%%%%%%%%%%%%%%%%%%%%%%%%%%%%%%%%%%%%%%%%%%%%%%%%%%%%%%%%%%%%%%%%%%%%%%%%%%%
%%%%%%%%%%%%%%%%%%%%%%%%%%%%%%%%%%%%%%%%%%%%%%%%%%%%%%%%%%%%%%%%%%%%%%%%%%%%%%%
\subsection{Covariance matrix}
\label{sec:cov}

The main subject of our analysis is the evolution of the covariance
matrix of this system, namely the two-point correlation functions of
the phase space variables. In order to develop a formalism that is
both able to describe the case of fixed and free b.c., we consider the
correlators of relative displacements and momentum, namely $\{\Delta
q_i = q_i - q_{i-1},p_i\}$.  More precisely, we study the covariance
matrix,
\begin{equation} \label{eq:covmat}
\mC = \left(
\begin{array}{cc}
  \mY       &   \mZ\\
  \mZ^\dag  &   \mV
\end{array}
\right) \ .
\end{equation}
where the correlation matrices $\mY$, $\mZ$ and $\mV$, of respective dimension
$(N-1)\!\times\!(N-1)$, $(N-1)\!\times\!N$ and $N\!\times\!N$ are defined as
\begin{equation} \label{eq:defcov}
\mY_{i,j} = \cor{\Delta q_i \Delta q_j} \ , \\
\mZ_{i,j}  =  \cor{\Delta q_i p_j} \ , \\
\mV_{i,j}  =  \cor{p_i p_j} \ , \\
\end{equation}
and  $\langle\cdot\rangle$ denotes  the average  over phase  space probability
distribution  function  $P$.  Note  that  $\mY$  and  $\mV$ are  symmetric  by
definition, while $\mZ$ has no definite symmetry.

In terms of the relative displacements, the boundary conditions are defined by
imposing
\begin{equation} \label{eq:free}
\Delta q_1 = 0 \qquad  \textrm{and} \qquad \Delta q_{N+1} = 0 \ ,
\end{equation}
for free BC and 
\begin{equation} \label{eq:fixed}
\Delta q_1 = q_1 \qquad  \textrm{and} \qquad \Delta q_{N+1} = -q_N \ ,
\end{equation}
for fixed BC.

Using $\{\Delta q_i\}$ instead of  $\{q_i\}$ is necessary since the average of
$q_i$ is  not well defined  when free BC  are considered. This is  at variance
with  \cite{LMMP09}, where  we  considered the  correlators  of the  variables
$\{q_i,p_i\}$. However, one  can recover the old formulation  by noticing that
there  exists  a  simple  mapping  with the  correlators  $\mathbf{U}_{i,j}  =
\cor{q_i  q_j}$, $\mathbf{Z}_{i,j}  = \cor{q_i  p_j}$ and  $\mathbf{V}_{i,j} =
\cor{p_i  p_j}$, studied  in \cite{LMMP09},  namely
\begin{equ} \label{eq:map}
\mY_{i,j} = \mU_{i,j} - \mU_{i,j+1} -  \mU_{i+1,j} +  \mU_{i+1,j+1} \ ,
  \quad \mZ_{i,j} =  \mathbf{Z}_{i,j} - \mathbf{Z}_{i-1,j} \ , \ \textrm{and}
  \quad  \mV_{i,j} = \mathbf{V}_{i,j} \ .
\end{equ}

The  evolution  equations  for  $\mC$  have  two  contributions:  the  dynamic
contribution    directly    obtained   from    the    equations   of    motion
\eref{eq:eqs-motion}, and  the stochastic contribution that  is evaluated upon
multiplying  \eref{eq:Lcoll} by  $x_ix_j$ and  thereby integrating  over phase
space.

We say that a given correlator in $\mC$  is in the {\it bulk} of the system if
the index of the momentum variable is in $[2,N-1]$ and the index of $\Delta q$
is  in $[3,N-1]$  for free  BC and  in $[2,N]$  for fixed  BC.   The evolution
equations $\mC$ are in the bulk
\begin{equa}[2]
& \dot  \mY_{i,j} &  \ = \  & \mZ_{j,i}-\mZ_{j,i-1}+\mZ_{i,j}-\mZ_{i,j-1}  \ ,  \\
&   \dot    \mZ_{i,j}    &   \    =    \   &    \mV_{i,j}-\mV_{i-1,j}+
  \omega^2\left(\mY_{i,j+1}-\mY_{i,j}\right)+
  \gamma\left(\mZ_{i,j+1}+\mZ_{i,j-1}-2\mZ_{i,j}\right) \ , \label{eq:Cdot}\\ 
& \dot \mV_{i,j} & \ =  \ & \omega^2\left(\mZ_{j+1,i}-\mZ_{j,i}+
  \mZ_{i+1,j}-\mZ_{i,j}\right)   + \g\mW_{i,j} \ ,
\end{equa}
where  the  $N\!\times\!N$  collision   matrix  $\mW$,  corresponding  to  the
contribution from  the stochastic noise,  depends on the distance  between the
evaluated indexes $i$ and $j$ and can be written in a compact form as
\begin{eqnarray} \label{wmat}
\fl \mW_{ij} = \ddelta_{i,j}\big[\ddelta_{i,j-1}\big(\ddelta_{i,N}\mV_{i+1,j} +
\ddelta_{j,1}\mV_{i,j-1}\big) +\ddelta_{i,j+1}\big(\ddelta_{i,1}\mV_{i-1,j}
+ \ddelta_{j,N}\mV_{i,j+1}\big)\big]  \nonumber \\ 
+ \delta_{i,j}\big(\ddelta_{i,N}\mV_{i+1,j+1}
+\ddelta_{j,1}\mV_{i-1,j-1}\big)
-\big(2\big(\ddelta_{i,j-1}+\ddelta_{i,j+1}-\delta_{i,j}\big)  \nonumber \\  
-\delta_{i,1}-\delta_{i,N}-\delta_{j,1}-\delta_{j,N} +
\delta_{i,1}\delta_{j,1} + \delta_{i,N}\delta_{j,N}\big)\mV_{i,j} \ ,
\end{eqnarray}
where $\delta_{i,j}$ is the Kronecker delta function and $\ddelta_{i,j} \equiv
1-\delta_{i,j}$.  $\mW_{ij}$ also  holds for the boundary terms,  and since it
deals with momentum variables only, it is independent of the specific BC.

%%%%%%%%%%%%%%%%%%%%%%%%%%%%%%%%%%%%%%%%%%%%%%%%%%%%%%%%%%%%%%%%%%%%%%%%%%%%%%%
%%%%%%%%%%%%%%%%%%%%%%%%%%%%%%%%%%%%%%%%%%%%%%%%%%%%%%%%%%%%%%%%%%%%%%%%%%%%%%%
\subsection{Boundary conditions}
\label{sec:boundaryeqs}

As  a  consequence  of  the   physical  boundary  conditions  imposed  on  the
oscillators of the  chain edges, the border terms  of the covariance matrices,
follow a  dynamics that is different  from \eref{eq:Cdot}.  The  BC affect the
phase  space   variables  $\Delta   q_i$,  according  to   \eref{eq:free}  and
\eref{eq:fixed},    and    $p_i$,    through    the    boundary    terms    in
\eref{wmat}. Furthermore,  the coupling between  the oscillators at  the edges
and   the    heat   baths   affects   the   evolution    of   their   momentum
\eref{eq:eqs-motion}. In  this section we  content ourselves writing  down the
equations of motion of the  border covariances corresponding to the {\it first
  column} and  {\it last  row} of  each matrix. The  equations for  the matrix
elements in  the last column  and first row  can be obtained  analogously. The
latter are not a mirror image of the former. Nevertheless, one can verify that
to leading order in the continuum limit, they lead to the same behaviour.

For fixed  BC the equations  of motion for  the border matrix elements  of the
first matrix column (index $1$ for both phase space variables) are
\begin{eqnarray}
\fl \dot{\mY}_{i,1} = \ddelta_{i,N+1}\mZ_{1,i} - \ddelta_{i,1}\mZ_{1,i-1} +
\mZ_{i,1}\ , \label{fixed-y-1:A1} \\  
\fl \dot{\mZ}_{i,1} = \ddelta_{i,N+1}\mV_{i,1} - \ddelta_{i,1}\mV_{i-1,1} +
  \w^2\big(\mY_{i,2}-\mY_{i,1}\big) + \g\big(\mZ_{i,2} - \mZ_{i,1}\big)
  -\l\mZ_{i,1}\ ,\label{fixed-y-1:A2}  \\
\fl \dot{\mV}_{i,1} = \w^2\big(\mZ_{i+1,1} - \mZ_{i,1} + \mZ_{2,i} -
  \mZ_{1,i}\big) -\l\big(1+\delta_{i,1}+\delta_{i,N}\big)\mV_{i,1} +
  \delta_{i,1}2\l\kB T_+ + \g\mW_{i,1}\ ,\label{fixed-y-1:A3}
\end{eqnarray}
where $1\le i \le N+1$ in \eref{fixed-y-1:A1}, \eref{fixed-y-1:A2} and $1\le i
\le N$  in \eref{fixed-y-1:A3}.   For fixed BC  the last matrix  row different
from zero corresponds to index $N+1$ for $\Delta q$ variables and to index $N$
for $p$ variables. The corresponding equations of motion are
\begin{eqnarray}
\fl \dot{\mY}_{N+1,j} = -\mZ_{j,N} + \ddelta_{j,N+1}\mZ_{N+1,j} -
\ddelta_{j,1}\mZ_{N+1,j-1}\ ,\label{fixed-y+1:A1} \\  
\fl \dot{\mZ}_{N+1,j} = -\mV_{N,j} +
  \w^2\big(\mY_{N+1,j+1}-\mY_{N+1,j}\big) + 
\g\big(\ddelta_{j,N}\mZ_{N+1,j+1} + \ddelta_{j,1}\mZ_{N+1,j-1} \nonumber\\
 - \big(\ddelta_{j,N}+\ddelta_{j,1}\big)\mZ_{N+1,j}\big)
 \ , \label{fixed-y+1:A2}\\  
\fl \dot{\mV}_{N,j} = \w^2\big(\mZ_{N+1,j}-\mZ_{N,j} +
\mZ_{j+1,N} -  \mZ_{j,N}\big)
-\l\big(1+\delta_{j,1}+\delta_{j,N}\big)\mV_{N,j}  \nonumber\\ +
  \delta_{j,N}2\l\kB T_- + \g\mW_{N,j}\ ,\label{fixed-y+1:A3}
\end{eqnarray}
where  $1\le  j  \le  N+1$  in  \eref{fixed-y+1:A1} and  $1\le  j  \le  N$  in
\eref{fixed-y+1:A2}, \eref{fixed-y+1:A3}.

For free  BC the first  matrix column (different  from zero) is index  $2$ for
$\Delta q$ variables  and $1$ for $p$ variables.  Consequently, the equations
of motion are
\begin{eqnarray}
\fl \dot{\mY}_{i,2}  =  \mZ_{2,i} - \mZ_{2,i-1} + \mZ_{i,2} -
\mZ_{i,1}\ ,\label{free-y-1:A1} \\  
\fl \dot{\mZ}_{i,1} = \mV_{i,1} - \mV_{i-1,1} + \w^2\mY_{i,2} + \g\big(\mZ_{i,2} -
\mZ_{i,1}\big) -\l\mZ_{i,1}\ , \label{free-y-1:A2} \\
\fl \dot{\mV}_{i,1} = \w^2\big(\ddelta_{i,N}\mZ_{i+1,1} - \ddelta_{i,1}\mZ_{i,1} +
\mZ_{2,i}\big)  -\l\big(1+\delta_{i,1}+\delta_{i,N}\big)\mV_{i,1}+
\delta_{i,1}2\l\kB T_+ +\g\mW_{i,1}\ , \label{free-y-1:A3} 
\end{eqnarray}
where $2\le i \le N$ in \eref{free-y-1:A1}, \eref{free-y-1:A2} and $1\le i \le
N$  in \eref{free-y-1:A3}.  For  the border  matrix elements  of the  last row
(index $N$ for both $\Delta q$ and $p$ variables), the equations of motion are
\begin{eqnarray}
\fl \dot{\mY}_{N,j}  =  \mZ_{j,N} - \mZ_{j,N-1} + \mZ_{N,j} -
\mZ_{N,j-1}\ ,\label{free-y+1:A1} \\  
\fl \dot{\mZ}_{N,j} = \mV_{N,j}-\mV_{N-1,j} +
  \w^2\big(\ddelta_{j,N}\mY_{N,j+1}-\ddelta_{j,1}\mY_{N,j}\big)
  -\l\big(\delta_{j,1}+\delta_{j,N}\big)\mZ_{N,j}\nonumber \\ 
+ \g\big(\ddelta_{j,N}\mZ_{N,j+1} + \ddelta_{j,1}\mZ_{N,j-1}
-\big(\ddelta_{j,1}+\ddelta_{j,N}\big)\mZ_{N,j}\big)\ ,\label{free-y+1:A2} \\
\fl \dot{\mV}_{N,j} = \w^2\big(-\mZ_{N,j} -
  \mZ_{j,N}+\ddelta_{j,N}\mZ_{j+1,N}\big)
  -\l\big(1+\delta_{j,1}+\delta_{j,N}\big)\mV_{N,j} \nonumber\\
+ \delta_{j,N}2\l\kB T_- +  \g\mW_{N,j} \ ,\label{free-y+1:A3}
\end{eqnarray}
where  $2\le  j   \le  N$  in  \eref{free-y+1:A1}  and  $1\le   j  \le  N$  in
\eref{free-y+1:A2}, \eref{free-y+1:A3}.

The equations presented  in this section (together with  the equations for the
border  elements of the  first row  and of  the last  column of  the matrices)
constitute the dynamic boundary conditions of \eref{eq:Cdot}.

%%%%%%%%%%%%%%%%%%%%%%%%%%%%%%%%%%%%%%%%%%%%%%%%%%%%%%%%%%%%%%%%%%%%%%%%%%%%%%%
%%%%%%%%%%%%%%%%%%%%%%%%%%%%%%%%%%%%%%%%%%%%%%%%%%%%%%%%%%%%%%%%%%%%%%%%%%%%%%%
\subsection{Change of variables}
\label{sec:change}

We  shall  see  in  the   following  sections  that  certain  combinations  of
covariances  appear naturally in the  evolution  equations.   It is  therefore
convenient to introduce some of these combinations as new variables. An
important example of such combinations is
\begin{equation} \label{Psi-def}
\psi_{i,j} = \mV_{i,j} - \omega^2\mY_{i,j} \ ,
\end{equation}
which  has a  precise  physical meaning:  its  diagonal elements  $\psi_{i,i}$
correspond to the local balance between kinetic and potential energies. If our
system satisfies the virial theorem, then  we should find that $\psi_{i,i} = 0$
for all $i$.  As we will discuss in the  following section \ref{sec:bbx}, this
is not always the  case. In the rest of the paper  we will substitute $\mY$ in
favour of $\psi$.

In \cite{LMMP09} we found that the stationary state solution of \eref{eq:Cdot}
obtained by taking  all time derivatives to zero,  implies that the covariance
matrix $\mathbf{Z}$ is antisymmetric.  From \eref{eq:map} it is clear that
the covariance $\mZ$ has no definite symmetry, not even  in the
stationary state.  Therefore, it is pertinent,  as we do in  the following, to
decompose $\mZ$ in its symmetric and antisymmetric components $\mZ^\pm$
\begin{equation} \label{Zsym}
\mZ_{i,j} = \mZ^+_{i,j} + \mZ^-_{i,j} \ , \quad \mathrm{with} \ \ \
\mZ^\pm_{i,j} = \frac{\mZ_{i,j} \pm \mZ_{j,i}}{2} \ .
\end{equation}

In these variables, the equations of motion \eref{eq:Cdot} are
\numparts
\begin{eqnarray} \label{eq:dynamics-1}
\fl \dot \psi_{i,j} = \omega^2\big[ -4\mZ^+_{i,j}+\mZ^+_{i,j+1} +
 \mZ^+_{i+1,j} + \mZ^+_{i-1,j} + \mZ^+_{i,j-1} -\mZ^-_{i,j+1} +
 \mZ^-_{i+1,j} - \mZ^-_{i-1,j} + \mZ^-_{i,j-1}\big] \nonumber\\
+ \gamma\mW_{i,j} \ , \label{eq:A}\\ 
\fl 2\dot \mZ^-_{i,j} = \gamma\phantom{^2}\big[\mZ^-_{i,j+1} +  \mZ^-_{i+1,j}
  +  \mZ^-_{i-1,j} + \mZ^-_{i,j-1} + \mZ^+_{i,j+1} -  \mZ^+_{i+1,j} -
  \mZ^+_{i-1,j} + \mZ^+_{i,j-1} -4\mZ^-_{i,j}\big]\nonumber\\ + \psi_{i+1,j} -
\psi_{i,j+1} + \mV_{i,j+1} - \mV_{i+1,j} -  \mV_{i-1,j} +   \mV_{i,j-1}
\ , \label{eq:B}\\    
\fl 2\dot \mZ^+_{i,j}  =  \gamma\phantom{^2}\big[\mZ^+_{i,j+1} +
  \mZ^+_{i+1,j} +  \mZ^+_{i-1,j} + \mZ^+_{i,j-1} + \mZ^-_{i,j+1} -
  \mZ^-_{i+1,j} -  \mZ^-_{i-1,j} + \mZ^-_{i,j-1}-4\mZ^+_{i,j}\big]
\nonumber\\ + 2\psi_{i,j} - \psi_{i+1,j} - \psi_{i,j+1} + \mV_{i,j+1} + 
 \mV_{i+1,j} - \mV_{i-1,j} -   \mV_{i,j-1}\ , \label{eq:C}\\    
\fl \dot \mV_{i,j} = \omega^2\left[-2\mZ^+_{i,j} + \mZ^+_{i,j+1} +
 \mZ^+_{i+1,j}  - \mZ^-_{i,j+1} + \mZ^-_{i+1,j}\right] + 
\gamma\mW_{i,j}\ , \label{eq:D} 
\end{eqnarray}
\endnumparts

Other useful combinations of covariances will be considered when needed.

%%%%%%%%%%%%%%%%%%%%%%%%%%%%%%%%%%%%%%%%%%%%%%%%%%%%%%%%%%%%%%%%%%%%%%%%%%%%%%%
%%%%%%%%%%%%%%%%%%%%%%%%%%%%%%%%%%%%%%%%%%%%%%%%%%%%%%%%%%%%%%%%%%%%%%%%%%%%%%%
\subsection{Perturbative expansion and continuum limit}
\label{sec:expansion}

Our   first  goal   is  to   transform   the  set   of  difference   equations
\eref{eq:A}-\eref{eq:D}  into  a set  of  partial  differential equations,  by
taking  a continuum  limit that  is appropriate  to our  problem.  We do this
by following a perturbative-like analysis and choose $\e =  1/\sqrt{N}$ as
perturbation parameter. In order to proceed forward we need first to attribute
the right order (in powers of $\e$) to the covariance matrix elements. In order
to keep the presentation as simple as possible, we proceed on the  basis of our
knowledge of the stationary solution \cite{LMMP09} and of the numerical solutions
discussed in \cite{partII}. 

In the stationary  state, $\mY_{i,j}$ and $\mV_{i,j}$ are  $\Or(\e)$, with the
exception of the  diagonal terms that are $\Or(1)$,  being proportional to the
mean  potential and  kinetic  energy,  respectively.  On  the  other hand  the
combination  in equation  \eref{Psi-def} is  $\Or(\e^2)$, indicating  that the
system is  locally at equilibrium. Moreover,  it can be shown  that the fields
$\mZ^\pm$   are   $x$-derivatives   of  $\mathbf{Z}$,   namely   $\mZ^+\propto
\mathbf{Z}_x$  and $\mZ^-\propto \mathbf{Z}_{xx}$.  In \cite{LMMP09}  we found
that off-diagonal terms of $\mathbf{Z}$ are $\Or(1)$, while along the diagonal
$\mathbf{Z}=0$ due,  to its antisymmetry. Since each  differentiation \wrt $x$
increases the  order by  $\e$ (see \eref{eq:xyderiv}  below), $\mZ^+$  must be
$\Or(\e)$ and $\mZ^-$ of $\Or(\e^2)$.
Altogether,
\begin{equation} \label{eq:order}
\begin{array}{rcl}
\psi_{i,j} \ & \equiv & \ \e^2\delta_{i,j}\Omega_{i} +
  \e^2\left(1-\delta_{i,j}\right)  \nPsi_{i,j} \ , \nonumber\\
\mZ^+_{i,j} \ & \equiv & \ \e\delta_{i,j}\nS^+_{i} +
 \e\left(1-\delta_{i,j}\right){\nZ}^+_{i,j} \ , \nonumber\\
\mZ^-_{i,j} \ & \equiv & \
  \e^2\left(1-\delta_{i,j}\right){\nZ}^-_{i,j} \ , \nonumber\\
\mV_{i,j}  \ & \equiv & \ \delta_{i,j}{T}_{i} +
  \e\left(1-\delta_{i,j}\right){\nV}_{i,j} \ . 
\end{array}
\end{equation}
These equations  fix also the  notation that  we will use  in the rest  of the
paper. Note that  in this notation, all the correlators  appearing at the \rhs
of the definitions above, namely $\Omega$, $\nPsi$, $\nS^+$, $\nZ^+$, $\nZ^-$,
$T$ and $\nV$,  are $\Or(1)$.  We refer the reader  to \cite{partII}, where we
give numerical  support for the validity of  \eref{eq:order}.  Furthermore, we
have found  in \cite{partII} that for  fixed BC, the  off-diagonal elements of
$\psi$ scale as  $\e^3$ for $\lambda = \omega$, and  as in \eref{eq:order} for
$\lambda \ne  \omega $.   In what follows  we will assume  \eref{eq:order} and
discuss  the modifications for the resonant  case  $\lambda  =  \omega$  where
appropriate.

The last step before proceeding  to the derivation of the partial differential
equations  consists in transforming  the discrete  indexes of  the correlators
into two suitable  continuous variables. We do this by introducing the variable
$y$ for the diagonal direction  and variable $x$ for the transversal direction
as (see {\em e.g.}, figure 1 in \cite{LMMP09}),
\begin{equation} \label{def:xy}
x \equiv (i-j)\e \ ; \qquad y \equiv \frac{(i+j)\e^2 -1}{1-|i-j|\e^2}\ .
\end{equation}
The  nonlinear definition  of $y$  is chosen  so that  its domain  $[-1,1]$ is
independent  of  $x$.  Nevertheless,  in the  limit  $N\rightarrow\infty$  the
effects of  the nonlinearities are localized  at the boundaries  of the domain
and  thus, do  not complicate  the study  of the  bulk  dynamics. Differential
changes in these variables can be written as
\begin{equation} \label{eq:xyderiv}
x' = x + f\e \ ; \qquad  y' = \frac{(i+j)\e^2-1+s\e^2}{1-|i-j|\e^2 -f\e^2} \ ,
\end{equation}
where the  integer {\it shift}  functions $f(\Delta i,\Delta j)  \equiv \Delta
i\!  -\!   \Delta j$  and $s(\Delta  i,\Delta j) \equiv  \Delta i+  \Delta j$,
($f,s:\mathbb{Z}^2\mapsto\mathbb{Z}$),  account for  the displacements  on the
discrete variables along the $x$ and $y$ directions respectively.

Using   this,  the   continuum  limit   of  any   covariance   matrix  element
$\mathbf{m}_{i+\Delta i,j+\Delta j}$ can be written up to $\Or(\e^2)$ as
\begin{equation}\label{rule}
\mathbf{m}_{i+\Delta i,j+\Delta j} = 
\mathbf{m}(x+f\e, y+\e^2(fy+s))\ , \quad\textrm{for} \ i \ge j \ ,\\
\end{equation}
for  the  continuous  correlator  function  $\mathbf{m}$.  Here  and  in  what
follows, we keep  the same notation for the  continuous functions derived from
the  matrix variables ({\it  e.g.}  $\mV_{i,j}(t)  \longrightarrow \mV(x,y,t)$
{\it etc.}).  If we  restrict to the matrix {\it lower triangle},  as we do in
\eref{rule} and in  the following, then in the  limit $N\rightarrow\infty$ the
variables $(x,y)$ belong to the domain
\begin{equation} \label{def:D}
\mathcal{D} \equiv \left\{(x,y) | x\in[0,\infty); \ y\in[-1,1] \right\}
\end{equation}
Note, for instance, that $x\!=\!const$  corresponds to moving along the diagonal
direction, ($x\!=\!0$  corresponding to the main diagonal).   For more details
we refer the reader to section 4.1 of \cite{LMMP09}.

%%%%%%%%%%%%%%%%%%%%%%%%%%%%%%%%%%%%%%%%%%%%%%%%%%%%%%%%%%%%%%%%%%%%%%%%%%%%%%%
%%%%%%%%%%%%%%%%%%%%%%%%%%%%%%%%%%%%%%%%%%%%%%%%%%%%%%%%%%%%%%%%%%%%%%%%%%%%%%%
\section{Main results}
\label{sec:results}

In  the next  section we  show  that, after  eliminating the  fast degrees  of
motion,  the  bulk  dynamics  reduces  to the  following  two  equations  (the
subscript denoting partial derivation)
\begin{eqnarray}
\dot{{\nZ}}^+ &  = \ & \e^2\left(
 \gamma{\nZ}^+_{xx} + 2{\nV}_{y} \right) \ , \label{slow-1}\\
\dot{{\nV}} &  = \ & \e^2\left(
\gamma  {\nV}_{xx} + 2\omega^2{\nZ}^+_{y} \right) \ ,\label{slow-2}
\end{eqnarray}
together  with the  conditions on  the  boundary of  the domain  \eref{def:D},
\begin{equ}
\delta\mathcal{D} \equiv \{x=0\} \cup \{y=\pm1\} \ .
\end{equ}
On $x=0$, we find that
\begin{eqnarray} \label{bbb-3}
\nV_x(0,y,t) = 0 \ , \\
\label{bbb-4}
\nZ^+_x(0,y,t) = -\frac{1}{\g}T_y \ , \\
\label{bbb-8}
\dot{T}(y,t) = 2\e^3\omega^2{\nZ}_y^+(0,y,t) \ .
\end{eqnarray}
This last  equation is accompanied  by its BC, namely  $T(\mp1,t)=T_\pm$.  The
peculiar  mathematical structure  of  the problem  should  be underlined:  the
boundary  evolution of  $\nZ$ is  determined dynamically  by  the differential
equations  \eref{bbb-4}   and  \eref{bbb-8}  which,  in   turn  determine  the
temperature field $T(y,t)$.

On $y=\pm 1$ the BC are particularly difficult because close to these
boundaries, the expansions \eref{eq:order} are no longer valid.  The
numerical solution in \cite{partII}, shows that a kind of ``boundary
layers'' (BL) exist, namely that at least some of the fields scale
differently in the regions of $\mathcal D$ lying within a distance
$\Or(\e)$ from $y=\pm 1$. In particular, in this region, $\psi$ is of
order $\Or(\e)$.  Recalling that $\psi$ is the difference between the
kinetic and potential energy, which away from the diagonal are
$\Or(\e)$, this implies that in the BL the relative difference between
potential and kinetic terms is of $\Or(1)$, which means that the
system is not in local thermal equilibrium.

The existence of a BL hinders us from finding an explicit exact
solution in the general case. This technical difficulty is irrelevant
in two cases: $a$) for fixed BC and $b$) for free BC at the
``resonant'' value of the bath coupling constant, $\lambda=\omega$. In
case ($a$), the boundary layer does not affect the relevant fields
and, as we show in section \ref{sec:bc}, it is sufficient to impose
\begin{equation}
\label{eq:fixedbc-3} \nZ^+(x,\pm 1,t) = 0 \qquad , \qquad \nV(x,\pm 1,t) = 0
\quad.
\end{equation}
This allows us  to find an explicit time-dependent  solution.  From a physical
point  of view,  case ($b$)  corresponds to  the only  situation in  which the
coupling  can  be  perfectly tuned  to  avoid  an  impedance mismatch  on  the
boundaries.  Unfortunately, we  were not able to find  an explicit solution in
this case.

Inspection of the  equations of motion reveals that  the temperature field $T$
is    the    slowest    variable    (it    evolves    on    a    time    scale
$O(\varepsilon^{-3})$). Since ${{\nZ}}^+$ and  ${{\nV}}$ relax on a time scale
$O(\varepsilon^{-2})$ they  can be  adiabatically eliminated by  setting their
time derivative equal to zero in equations \eref{slow-1} and \eref{slow-2}. By
further eliminating the field $\nV$, we obtain the fourth-order equation
 \begin{equation}
\label{eq:old}
  \gamma^2{\nZ}^+_{xxxx} - 4\omega^2{\nZ}^+_{yy} = 0 \ ,
\end{equation}
that  is formally  equal to  the equation  solved in  \cite{LMMP09},  the main
difference being that here $\nZ^+$  is time dependent.  The dynamical equation
is obtained by first determining the stationary solution of \eref{eq:old} with
the  appropriate  BC and  then  using \eref{bbb-4}  to  express  $\nZ^+$ as  a
function  of  $T$  and  replacing  the  result  into  \eref{bbb-8}.   This  is
accomplished by considering the Fourier expansion
\begin{equation}
T(y,t) = 
T_s(y) + \sum_{n=1}^\infty \mathcal{T}_n(t) 
\sin\left[\frac{n\pi}{2}\left(y+1\right)\right] \, ,
\label{expt}
\end{equation}
with $T_s$ being the stationary solution of $T$, as given by formulas (18) and
(19)  of   \cite{LMMP09}.   In  section  \ref{sec:relax}  we   show  that  the
coefficients  $\mathcal{T}_n(t)$ obey  the  linear equation  \eref{eq:eraora}.
The associated eigenvalues,  that must be computed numerically  as the problem
is not exactly  diagonal, uniquely determine the relaxation to the steady state
for  any assigned initial  condition.  They  are found  to be  proportional to
$-(k/N)^{3/2}$ (for  $k$ being a  positive integer labeling  the eigenvalues).
This  is reminiscent  of the  spectrum of  the eigenvalues  of  the fractional
Laplacian  $\nabla^{3/2}$,   thus  suggesting   that  the  evolution   of  the
temperature  field  is ruled  by some underlying fractional  diffusion
equation on hydrodynamic scales \cite{JKO08,ZRK07}.

%%%%%%%%%%%%%%%%%%%%%%%%%%%%%%%%%%%%%%%%%%%%%%%%%%%%%%%%%%%%%%%%%%%%%%%%%%%%%%%
%%%%%%%%%%%%%%%%%%%%%%%%%%%%%%%%%%%%%%%%%%%%%%%%%%%%%%%%%%%%%%%%%%%%%%%%%%%%%%%
\section{Dynamical equations}
\label{sec:dynamics}

We first focus on the dynamics in the ``bulk'' of system, namely the
interior of $\mathcal D$. In \ref{ap:diff-bulk} we derive the set of
coupled partial differential equations
\eref{bulk-diff:A}-\eref{bulk-diff:D}, for the covariances $\psi$,
$\mZ^-$, $\mZ^+$ and $\mV$.  Furthermore, using \eref{eq:order} the
equations with their explicit order in $\e$ are rewritten as \numparts
\begin{eqnarray}
  \dot{{\Psi} } = 
 2\e\left(
 2\omega^2{\nZ}^-_x + \omega^2{\nZ}^+_{xx} + \gamma
 {\nV}_{xx} \right)
 + 4\e^2\omega^2y{\nZ}^-_{y} \ ,  \label{bulk-diff:A2}\\
  \dot{{\nZ}}^- = 
 \e{\Psi}_x +
 \e^2\left(\gamma{\nZ}^-_{xx} + y{\Psi}_{y}\right)
 \ ,  \label{bulk-diff:B2}\\
  \dot{{\nZ}}^+ = 
 \e^2\left(\gamma{\nZ}^+_{xx}
 + 2{\nV}_{y}
 \right) - \e^3\left( \frac{1}{2}{\Psi}_{xx}
 + {\Psi}_{y} \right)
 \ ,  \label{bulk-diff:C2}\\
  \dot{{\nV}} = 
 \e^2\left(
2\omega^2{\nZ}^-_x
 +\omega^2{\nZ}^+_{xx}
 +2\gamma  {\nV}_{xx}
 + 2\omega^2{\nZ}^+_{y}
\right)
 + 2\e^3\omega^2y{\nZ}^-_{y}
 \ . \label{bulk-diff:D2}
 \end{eqnarray}
\endnumparts  The four  variables  can be  split  into a  pair  of {\em  fast}
(${\Psi}$ and  ${\nZ}^-$) and {\em  slow} (${\nZ}^+$ and ${\nV}$)  ones, which
evolve on  time scales of  order $\e^{-1}$ and $\e^{-2}$,  respectively.  Upon
substituting   the  $x$-derivative  of   \eref{bulk-diff:A2}  into   the  time
derivative of \eref{bulk-diff:B2} we find
\begin{eqnarray} \label{fast-1}
\fl \ddot{{\nZ}}^- - \ 4\e^2\omega^2{\nZ}^-_{xx} -
2\e^2\left(\omega^2{\nZ}^+_{xxx} + \gamma {\nV}_{xxx} \right) =   \nonumber\\ 
\e^2\gamma \dot \nZ^-_{xx} + 8\e^3\omega^2 y\nZ^-_{xy} + 2 \e^3
y\left(\omega^2 \nZ^+_{xxy} + \nV_{xxxy} \right)  \ ,
\end{eqnarray}
where we have retained  only terms up to order $\e^3$.  The  terms on the \rhs
do not affect the final solution but must be taken into account to justify the
adiabatic elimination. In fact, scaling the time by $\e$, we see that they are
$o(\e^3)$ and could, in principle, be neglected.  However, if we do so, we are
left with a non-dissipative wave equation (with source terms) for $\nZ^-$, that
cannot account  for the convergence  towards the steady state.   Therefore, to
study  the evolution  of the  fast variables,  we are  obliged to  include the
higher  order terms  (losses are  actually provided  by the  $\dot \nZ^-_{xx}$
term, the other  being perturbations of the source  term).  After this remark,
we  are  authorized  to  adiabatically  eliminate $\nZ^-$  and  $\Psi$.   From
(\ref{bulk-diff:A2}, \ref{bulk-diff:B2}) we obtain to leading order \numparts
\begin{eqnarray}
 {\nZ}^-_x &=& -\frac{1}{2}{\nZ}^+_{xx} + \frac{\gamma}{2\omega^2}{\nV}_{xx}
 \ , \label{bulk-adiab1}\\
  \Psi_x &=& 0 \ . \label{bulk-adiab2}
 \end{eqnarray}
\endnumparts Note that  $\Psi$ is constant moving away  from the diagonal.  We
now turn  our attention  to the  slow variables and  substitute the  above two
equations into (\ref{bulk-diff:C2},  \ref{bulk-diff:D2}). To leading order, we
finally obtain \eref{slow-1} and \eref{slow-2}.

At this point,  it is useful to illustrate some  features of \eref{slow-1} and
\eref{slow-2}. The symmetry  of these equations suggests to  introduce the new
variables  $\mathcal{Q^{(\pm)}}   =  \omega{\nZ}^+  \pm   {\nV}$,  that  allow
decoupling the system of equations into
\begin{equation} \label{slow-3}
 \dot{\mathcal{Q}}^{(\pm)} = \e^2\left(\gamma\mathcal{Q}^{(\pm)}_{xx} \pm
 2\omega\mathcal{Q}^{(\pm)}_{y}\right)   \ .
\end{equation}
The  first  term  on  the  \rhs  describes  a  transversal  diffusion  process
characterized by the diffusion constant $\e^2\gamma$; the second term accounts
for  a longitudinal  right/left sound-wave  propagation, depending  on whether
$\mathcal{Q}^-$/$\mathcal{Q}^+$ is  considered.  Leaving aside  for the moment
the issue of BC,  by absorbing the order $\e^2$ in the  time variable, we look
for solutions of the form
\begin{equation} \label{sa-1}
\mathcal{Q}^+(x,y,t) = P(x,t)\cos K_y y + Q(x,t) \sin K_y y \ ,
\end{equation}
we consider only $\mathcal{Q}^+$ as  the equation for $\mathcal{Q}^-$ leads to
the  same dispersion  relation).  By  substituting this  in  \eref{slow-3} and
separating the independent terms, we obtain
\begin{eqnarray} \label{sa-2}
\dot{P} = \g P_{xx} + 2\w K_y Q \ , \\
\dot{Q} = \g Q_{xx} - 2\w K_y P \ .
\end{eqnarray}
By then assuming the following form of the solution,
\begin{eqnarray} \label{sa-3}
P(x,t) &=& a(t)\cos K_x x + b(t) \sin K_x x \ , \\
Q(x,t) &=& c(t)\cos K_x x + d(t) \sin K_x x \ ,
\end{eqnarray}
we find that  $a(t)$ and $c(t)$ satisfy a system  of two ordinary differential
equations,
\begin{eqnarray} \label{sa-4}
\dot{a} &=& -\g K_x^2 a - 2\w K_y c \ , \\ 
\dot{c} &=& -\g K_x^2 c + 2\w K_y a \ , 
\end{eqnarray}
($b$  and $d$  do not  add any  information as  they satisfy  the same  set of
equations). Looking for  solutions of the form $a(t)  = \tilde{a}\exp(\mu t)$,
$c(t) =  \tilde{c}\exp(\mu t)$, the  resulting eigenvalue equation  yields two
degenerate branches for the dispersion relations,
\begin{equation} \label{sa-7}
\mu = -\g K_x^2  \pm 2\w i K_y \ ,
\end{equation}
where $i$ denotes the imaginary  unit.

Taking   $\mu=0$,  we  recover   the  eigenvalue   of  the   stationary  state
\cite{LMMP09}.   Most important,  the real  part  of $\mu$  is negative,  thus
ensuring the  stability of the  stationary state.  By reintroducing  the $\e^2$
factor in the  time units, we can thus conclude that  modes characterized by a
$K_x$ of $\Or(1)$  relax on a time scale of order  $\e^{-2}$.  As discussed in
\cite{partII},  these are  the modes  that mostly  contribute to  the relevant
nonzero off-diagonal correlations.  However, $K_x$ can, by construction, be as
small as  $\e^{-1}$.  As a  result the slowest  relaxation times that  one can
observe are of the order of  $N^2$.  This is indeed confirmed by the numerical
calculation of the spectrum of the evolution operator \cite{partII}.

%%%%%%%%%%%%%%%%%%%%%%%%%%%%%%%%%%%%%%%%%%%%%%%%%%%%%%%%%%%%%%%%%%%%%%%%%%%%%%%
\section{Boundary conditions}
\label{sec:bc}

A  complete solution  of the  dynamical problem  requires solving  the  set of
partial differential  equations \eref{bulk-diff:A2}-\eref{bulk-diff:D2} on the
whole domain $\mathcal{D}$ \eref{def:D},  including its boundary at all times.
In a general  context, the difficulty of obtaining a  full solution depends on
the  constraints  along  $\delta\mathcal{D}$.   If they  amount  to  algebraic
conditions or if the boundary dynamics  is faster than the bulk dynamics, then
one deals with standard type  of {\it static} BC.  In the opposite case, it is
the  bulk dynamics  that  can  be adiabatically  eliminated  and the  relevant
(long-term)  evolution   would  be  controlled  by  what   happens  along  the
boundaries.

In the first section we study the dynamics of the covariance elements close to
the diagonal  and derive a  differential equation describing the  evolution of
the  temperature profile.  Later, in  section \ref{sec:bby}  we study  how the
physical BC  determine the  dynamics of the  correlators along  the boundaries
$y=\pm1$.

%%%%%%%%%%%%%%%%%%%%%%%%%%%%%%%%%%%%%%%%%%%%%%%%%%%%%%%%%%%%%%%%%%%%%%%%%%%%%%%
\subsection{Boundary conditions along $x\!=0$}
\label{sec:bbx}

The BC along the diagonal are not connected with the physical BC of the chain,
but rather with  the symmetry of the various  matrices. In \ref{ap:diff-diag},
we  derive  the  differential   equations  describing  the  evolution  of  the
covariances     at     $x=0$,    (\ref{diag-diff-A}-\ref{diag-diff-D})     and
(\ref{ldiag-diff-A}-\ref{ldiag-diff-D}).  As  the diagonal terms  $\Omega$ and
$\nS^+$  appear  only  as   differences  with  their  respective  off-diagonal
counterparts, it  is convenient to introduce  ${\delta\Psi} = {\Psi}-{\Omega}$
and  $\e\delta\nZ^+(y,t) = {\nZ^+}(0,y,t)-{\nS^+}(y,t)$,  where we  benefit of
the numerical  investigations to anticipate  that the latter difference  is of
higher order than the single  addenda.\footnote{As a matter of fact, assigning
  to $\delta\nZ^+$ the  same scaling as its addenda  leads to {\it unphysical}
  super fast evolution.}

By introducing  ${\delta\Psi}$ and ${\delta\nZ^+}$  and using \eref{eq:order},
we obtain the  following set of partial differential  equations describing the
evolution at $x=0$: \numparts
\begin{eqnarray}
\dot{{\delta\Psi}} = 
-2\w^2\big({3\delta\nZ}^++{\nZ}^-+2{\nZ}^+_x\big)+2\g{\nV}_x+
\e\big(-4\w^2y{\nZ}^+_{y}+ \g \KV \big)
\ ,\label{bb-A}\\
\dot{{\Psi}} = 
2\w^2\big({{\nZ}^--\delta\nZ}^+\big)+2\gamma{\nV}_x +
\e\big(2\w^2\big({\nZ}^+_{xx}+2{\nZ}^-_x\big)+ \g\KV \big)
\ , \label{bb-B}\\  
\dot{{\nZ}}^- = 
\frac{{\delta\Psi}}{2}-\gamma{\nZ}^-+{T}_{y} +
\e\big({\Psi}_x-{\nV}_{y}\big)
\ ,\label{bb-C}\\ 
\dot{{\delta\nZ}}^+  = 
 \frac{3}{2}{\delta\Psi}-\gamma\big(3{\delta\nZ}^++2{\nZ}^+_x\big)
+{T}_{y} + \e\big({\Psi}_x-2\gamma y{\nZ}^+_{y}-{\nV}_{y}\big) 
\ ,\label{bb-D}\\
\dot{{\nZ}}^+ = 
\e\big(\frac{1}{2}{\delta\Psi} -\gamma{\delta\nZ}^++{T}_{y}\big)+
\e^2\big({\nV}_{y}+\gamma{\nZ}^+_{xx}\big)
\ , \label{bb-E}\\ 
 \dot{{\nV}} = 
\e\big(\omega^2\big({\nZ}^--{\delta\nZ}^+\big)+2\gamma{\nV}_x\big) + 
\e^2\big(\omega^2\big({\nZ}^+_{xx}+2{\nZ}^-_x+2{\nZ}^+_{y}\big) +
 \gamma\KV\big)
\ , \label{bb-F} \\
 \dot{{T}} = 
2\e^2\omega^2\big({\delta\nZ}^++{\nZ}^-+{\nZ}^+_x\big)+
\e^3\omega^2\big({\nZ}^+_{xx}+2{\nZ}^-_x+2\big(y+1\big){\nZ}^+_{y}\big)
\ ,\label{bb-G}
\end{eqnarray}
\endnumparts where we  have considered the first two  leading contributions to
the  evolution of  each variable  and, for  the sake  of compactness,  we have
introduced
\begin{equation} \label{eq:K}
\KV = 3{\nV}_{xx}+2y{\nV}_{y} \ .
\end{equation}
Like in the bulk, the  variables evolve over manifestly different time scales,
and the  temperature field  $T$ is  the slowest one.  Therefore we  proceed by
adiabatically eliminating the other variables, starting from the fastest ones.
By setting the first three time derivatives equal to zero and considering only
the leading terms, equations  \eref{bb-A}, \eref{bb-B} and \eref{bb-C} lead to
\numparts
\begin{eqnarray} 
\delta\Psi = -\left(\g\nZ^+_x + \frac{\g^2}{\w^2}\nV_x  + 2T_y\right)
\ , \label{bbb-1a}\\
\nZ^- = -\frac{1}{2}\left(\frac{\g}{\w^2}\nV_x + \nZ^+_x\right)
\ , \label{bbb-1b}\\ 
\delta\nZ^+ = \frac{1}{2}\left(\frac{\g}{\w^2}\nV_x - \nZ^+_x\right)
\ . \label{bbb-1c} 
\end{eqnarray}
\endnumparts Moreover, the stationary solution of \eref{bb-D} yields
\begin{equation} \label{bbb-2}
\g\nZ^+_x + \frac{3\g^2}{2\w^2}\nV_x + T_y = 0 \ .
\end{equation}
By now inserting \eref{bbb-1a} and \eref{bbb-1c} into \eref{bb-E}, and setting
the  time derivative $\dot{{\nZ}}^+$  equal to  zero, we  obtain \eref{bbb-3}.
This equation has an obvious  meaning: $\nV$ is symmetric by definition across
the diagonal, so that we naturally expect $\nV$ to have be maximal for $x=0$.

Furthermore,  by using  \eref{bbb-3}  in \eref{bbb-2},  we  obtain the  second
relevant  constraint \eref{bbb-4}. From  \eref{bbb-1b} and  \eref{bbb-1c}, one
can easily find that $\nZ^-=\delta\nZ^+$.  By then going back to \eref{bbb-1a}
\eref{bbb-1b} and  \eref{bbb-1c}, and  using the constraints  \eref{bbb-3} and
\eref{bbb-4}), we obtain
\begin{equation} \label{bbb-5}
\delta\nZ^+ =
{\nZ}^- = \frac{1}{2\gamma}{T}_y \qquad , \qquad {\delta\Psi} = -{T}_y \ .
\end{equation}
Altogether, once  the five conditions contained  in \eref{bbb-3}, \eref{bbb-4}
and \eref{bbb-5} are satisfied, it turns out that the derivatives of all seven
boundary  variables  (\ref{bb-A}-\ref{bb-G})  are  equal to  zero  to  leading
order. In particular we see that the
variable $\Psi$ remains  undetermined, but this is not a problem,  as it is of
higher order in $\e$ and does not contribute to the leading-order evolution of
the physically relevant variables.  Additionally, with the exception of $\nV$,
all the variables are expressed as  a function of the temperature profile $T$,
whose  evolution must  be determined  if we  want to  find a  closed solution.

It  turns out  that the  leading contribution  of $\Or(\e^2)$  to $\dot  T$ is
zero. Therefore, it  is necessary to go one order  further in the perturbative
analysis.  This can be easily done  by noting that the leading contribution to
$\dot{T}$ is  equal to that  of $\dot{\Omega}=\dot{{\Psi}}-\dot{{\delta\Psi}}$
(see \ref{diag-diff-A}). Subtracting  \eref{bb-B} from \eref{bb-A} and setting
the time derivatives equal to zero we find that, up to $\Or(\e^2)$,
\begin{equation} \label{bbb-7}
2\omega^2\left({\delta\nZ}^++{\nZ}^-+{\nZ}^+_x\right) = 
-\e\omega^2\left({\nZ}_{xx}^++2{\nZ}_x^-+2y{\nZ}_y^+\right) \ .
\end{equation}
By inserting this  into \eref{bb-G} and retaining terms  up to $\Or(\e^3)$, we
obtain equation  \eref{bbb-8}.  By recalling that  ${\nZ}_y^+$ is proportional
to  the divergence  of the  heat flux  along the  diagonal, we  recognize that
\eref{bbb-8} is nothing  but the continuity equation for  the energy and could
have been  derived simply  on the basis  of physical arguments.   However, the
relaxation  of the  temperature  profile towards  the  stationary state,
occurs  on  a  time scale  that  is  $\Or(\e^{-3})$,  {\em i.e.},  for  $t\sim
N^{3/2}$.  As a consequence, we can  conclude that the bulk dynamics is faster
than  that occurring  along the  diagonal and  can, thereby,  be adiabatically
eliminated as anticipated in section~\ref{sec:results}.

Finally, in order  to complete the treatment, we  must complement \eref{bbb-8}
with  its physical  BC,  as  it is  a  (one-dimensional) partial  differential
equation.  Without  the need  of a formal  treatment, it is  easily understood
that  these  BC  are simply  $T(\pm1,t)=T_\mp$.   As  a  matter of  fact,  the
relaxation  on  the  boundaries  occurs  on  a  finite  time  scale  ($\approx
1/\lambda$), i.e.   it is  basically instantaneous with  respect to  the above
mentioned time scales.

%%%%%%%%%%%%%%%%%%%%%%%%%%%%%%%%%%%%%%%%%%%%%%%%%%%%%%%%%%%%%%%%%%%%%%%%%%%%%%%
\subsection{Boundary conditions along $y=\pm1$}
\label{sec:bby}

In  this  section  we  analyse  the  conditions  that  the  physical  boundary
conditions, either  fixed or  free, impose on  the covariances.  At  the chain
edges,  where the  system is  directly coupled  to the  heat  baths stochastic
evolution,  the dynamics  is  different from that in the bulk: on the one hand
the deterministic restoring force is not counterbalanced by the boundary and on
the other  hand, the stochastic collision  for the edge oscillators  is also a
``one-sided'' process.  Consequently, we  introduce new auxiliary variables to
distinguish the boundary dynamics  from its bulk counterpart.  More precisely,
we define $\phi_{i,1}  \equiv \psi_{i,1}$, $\z^\pm_{i,1} \equiv \mZ^\pm_{i,1}$
and  $\nu_{i,1}  \equiv \mV_{i,1}$.

In   \ref{ap:diff-border}  we  derive   the  partial   differential  equations
describing the dynamics of the covariance at $y\!=\!-1$ for fixed and free BC.
However, in this  case we are not entitled to use  \eref{eq:order} in order to
assign the  correct order  in $\e$  of the covariance  variables.  As  we have
discussed in Section \ref{sec:results}, there exist a boundary layer, namely a
region around $y=\pm1$ of size  $\e^{-1}$, where the scaling of the covariance
matrix on $\e$ differs from  \eref{eq:order}. This BL has been further studied
numerically in \cite{partII}. It is important  to note that if we insist using
\eref{eq:order} then mathematical consistency  requires that the order of {\it
  e.g.},  $\psi$ is  $\Or(\e)$  and not  $\Or(\e^2)$,  which is  also what  we
numerically  observe   in  \cite{partII}.   However,   \eref{eq:order}  cannot
differentiate the scalings in the BL  from those in the bulk. Consequently, in
this Section we do not  use the expansions \eref{eq:order} and limit ourselves
to extract some physical  information from the leading contribution determined
only by the  differential structure of the equations. First  we focus on fixed
BC.

For fixed  BC we end up  with four equations  for the bulk variables  and four
equations for the  boundary variables, \eref{fbby-3:A}-\eref{fbby-3:H}.  Since
all equations  evolve on  time scales of  $\Or(1)$, they can  be adiabatically
eliminated from the bulk dynamics, which  is at least $\Or(\e)$. By taking all
time derivatives to  zero and solving the resulting set  of relations, we find
that  in  the  stationary  state  all boundary  variables  coincide  with  the
respective bulk counterparts,
\begin{equation} \label{eq:fixedbc-1}
\z^+ = \mZ^+ \ , \quad
\z^- = \mZ^- \ , \quad
\nu = \mV \ \ \textrm{and} \quad
\phi = \psi \ , 
\end{equation}
and that
\begin{eqnarray} 
\mV(x,-1,t) = 0 \ , \label{eq:fixedbc-2a} \\
\mZ^+(x,-1,t) = -\mZ^-(x,-1,t)\label{eq:fixedbc-2b} \ .
\end{eqnarray}
It is  interesting to point  out that these  relations are independent  of the
parameters  of  the  system   and  that  \eref{eq:fixedbc-1}  imply  that  the
bulk variables are continuous at the boundaries. Recalling that in
\eref{eq:fixedbc-2b} $\mZ^-$ is
of  higher  order  than  its  symmetric  counterpart  $\mZ^+$,  we  do  obtain
\eref{eq:fixedbc-3}.  It  is interesting to  note that the  indetermination of
$\psi(x,-1)$  and $\mZ^-(x,-1)$, on  which the  effects of  the BL  are mostly
observed,  does  not  affect  the  leading order  dynamical  solution  of  the
physically relevant  fields. Therefore,  for fixed BC,  the existence of  a BL
does not impede  us from using the boundary  relations \eref{eq:fixedbc-3} and
determining the evolution of $\mV$ and $\mZ^+$.

We now turn our attention to free BC. As seen in \ref{ap:free}, in this case,
only two  auxiliary variables along the boundaries are necessary,
$\z_{i,1}\equiv\mZ_{i,1}$  corresponding to  the non  symmetrizable  term, and
$\nu_{i,1}\equiv\mV_{i,1}$.   From \eref{bby-3:A}-\eref{bby-3:D} we  find that
when the three conditions
\begin{equation}
\nu= \mV \ , \quad \tilde{\z} = \mZ^+ \ \textrm{and} \quad \mZ^- = 0 
\end{equation}
are satisfied, the four time derivatives  are equal to zero (to leading order),
thus  satisfying the  stationary  state solution.   By  using these  relations
in~\eref{bby-3:E}-\eref{bby-3:F},   we    obtain   the   relevant   mathematical
conditions,
\begin{eqnarray} 
\psi(x,-1,t) = \mV(x,-1,t) -\l\mZ^+(x,-1,t) \ , \label{eq:freebc-1a}\\
\w^2\mZ^+(x,-1,t) = \l\mV(x,-1,t) \label{eq:freebc-1b} \ .
\end{eqnarray}
There    are    two    main    differences    between    \eref{eq:fixedbc-2a},
\eref{eq:fixedbc-2b}  and  the  equations above:  first,  \eref{eq:freebc-1a},
\eref{eq:freebc-1b}  depend  on  the   variable  $\psi$,  which,  as  we  have
discussed,  it  is  the variable  whose  behaviour  is  most affected  by  the
BL.  Second and  more  important, the  free  BC relations  now  depend on  the
parameters $\lambda$ and $\omega$.

By combining \eref{eq:freebc-1a} and \eref{eq:freebc-1b} we find that
\begin{equation} \label{eq:freebc-2}
(\w^2-\l^2) \mZ^+(x,-1,t) =  \l \psi(x,-1,t)\ .
\end{equation}
If $\w =\l$,  then $\psi(x,-1)  = 0$  (or at least $\Or(\e^2)$), consistently
with our expectations from the bulk dynamics. In this resonant case the
boundary condition reduces to $\w\nZ^+(x,-1,t) = \nV(x,-1,t)$. Though simple,
we have not found a way to derive an explicit solution of the bulk equation
which satisfies this constraint.

In  the non  resonant  regime $\w^2\ne\l^2$,  \eref{eq:freebc-2} implies  that
$\mZ^+$ and  $\psi$ are of the  same order along the  boundary.  The numerical
studies presented in \cite{partII} confirm this prediction, but show also that
this  is because $\psi$  is of  $\Or(\e)$.  Such  an observation  is seemingly
inconsistent  with the bulk  analysis which  predicts $\psi$  to be  of higher
order.  As said, the only way to solve the paradox is by invoking the presence
of a BL connecting the two  different scaling regimes. 

Summarizing this  section, the  BC for $y=\pm  1$ reveal a  crucial difference
between fixed  BC and  free BC.  In  the former  case they are  independent of
parameters of the system, which in turns imply that the asymptotic profile, as
well as the leading term of the  heat flux are {\it universal}.  In the latter
case  both quantities  depend on  the coupling  strength with  the  heat baths
$\l$. In this  very atypical situation the contact with the  baths may lead to
measurable macroscopic effects.  Note in  particular, that the heat flux for a
system with anomalous thermal conductivity,  like the one that concerns us, in
general may  depend on the type  of boundary conditions, even  in the infinite
volume limit.

%%%%%%%%%%%%%%%%%%%%%%%%%%%%%%%%%%%%%%%%%%%%%%%%%%%%%%%%%%%%%%%%%%%%%%%%%%%%%%%
\section{Dynamics of the temperature field}
\label{sec:relax}

The  evolution  of the  temperature  profile  is  determined by  \eref{bbb-8},
subjected to conditions \eref{bbb-3}  and \eref{bbb-4}.  Since the temperature
field evolves on  a slower time scale, the bulk  dynamics can be adiabatically
eliminated. Therefore, it suffices to solve \eref{eq:old} for $\nZ^+$ and plug
its solution  into \eref{bbb-8}.  As discussed above,  an explicit calculation
is   feasible   only  for   fixed   BC,   where   the  first   of   conditions
\eref{eq:fixedbc-3} implies  that we can, following  \cite{LMMP09}, expand the
time-dependent solution of $\nZ^+$ as
\begin{equation} \label{eq:ansatz-1}
\nZ^+(x,y,t) = \sum_n B_n(x,t)
\sin\left[\frac{n\pi}{2}\left(y+1\right)\right] \ .
\end{equation}
The Fourier coefficients satisfy the ordinary differential equation
\begin{equation} \label{eq:aux-1}
\frac{\partial^4 B_n}{\partial x^4} =
-\left(\frac{n\pi\omega}{\gamma}\right)^2 B_n \ ,
\end{equation}
whose explicit solution yields
\begin{equation} \label{eq:aux-3}
B_n(x,t) = A_n(t)  \exp(-\an x) \sin(\an x)
+A_n'(t)  \exp(-\an x) \cos(\an x)  \ ,
\end{equation}
where  $\an  \equiv  \sqrt{n\pi\omega/2\gamma}$  and  we  have  discarded  the
components  which   diverge  for  $x  \to  \infty$.   By  differentiating  the
equilibrium solution of \eref{slow-1} with respect to $x$, we realize that the
condition \eref{bbb-3} is equivalent to $\nZ^+_{xxx}(0,y,t)=0$, which in turns
implies that $A_n=-A_n'$.

If one  is interested only  in the stationary solution,  equation \eref{bbb-8}
implies  that $\nZ^+(0,y)= \mathrm{constant}$,  namely that  the heat  flux is
constant  along the  chain.  This  condition transforms  itself  into distinct
equations for  the coefficients $\{A_n\}$,  which can therefore  be determined
(apart  from   a  multiplicative  factor).   Afterwards,  with   the  help  of
\eref{bbb-4}  we  can  determine  $T_s(y)$.  The  unknown  multiplicative  and
additive factors are eventually removed by imposing $T_s(\mp 1)=T_\pm$.  We do
not  report these  calculations  as  they would  closely  follow what  already
reported in \cite{LMMP09}.

Here we  wish to solve the  dynamical problem, particularly  for the
temperature field $T(y,t)$. Let  us consider its Fourier expansion \eref{expt}
where we  have only included the terms that are appropriate for
fixed  BC.    To  write  down  closed  equations   for  the  coefficients
$\mathcal{T}_n$,  we must  face the  problem that  the two  sides  of equation
\eref{bbb-8} are  expanded in a different  set of functions,  namely sines and
cosines, respectively  and the  problem is therefore  not diagonal.   By using
vector  notations with  an  obvious meaning  of  the symbols,  we obtain  from
\eref{bbb-4} and \eref{bbb-8},
\begin{eqnarray}
\label{eq:coeff}
\mathcal{A} &=& \frac{1}{2\gamma} {\bf D}{\bf R} \mathcal{T}   \\
\dot{\mathcal{T}} &=& 2\e^3 \omega^2 {\bf R}\mathcal{A}
\end{eqnarray}
where 
\begin{equation}
\label{eq:cases}
R_{n,k} = \left \{
\begin{array}{ll}
2k^2/(k^2-n^2) & \mathrm{for} \quad k+n \quad \mathrm{odd} \\
 0 & \mathrm{otherwise}
\end{array}
\right.
\end{equation}
and $D_{n,m}=\delta_{nm}/\an$. 
We can thus write a closed equation for $\mathcal{T}$,
\begin{equation}
\label{eq:eraora}
 \dot{\mathcal{T}}  = \frac{\e^3 \omega^2}{\gamma} {\bf RDR} \mathcal{T}
\end{equation}

\begin{figure}[!ht]
\begin{center}
\includegraphics[scale=0.55]{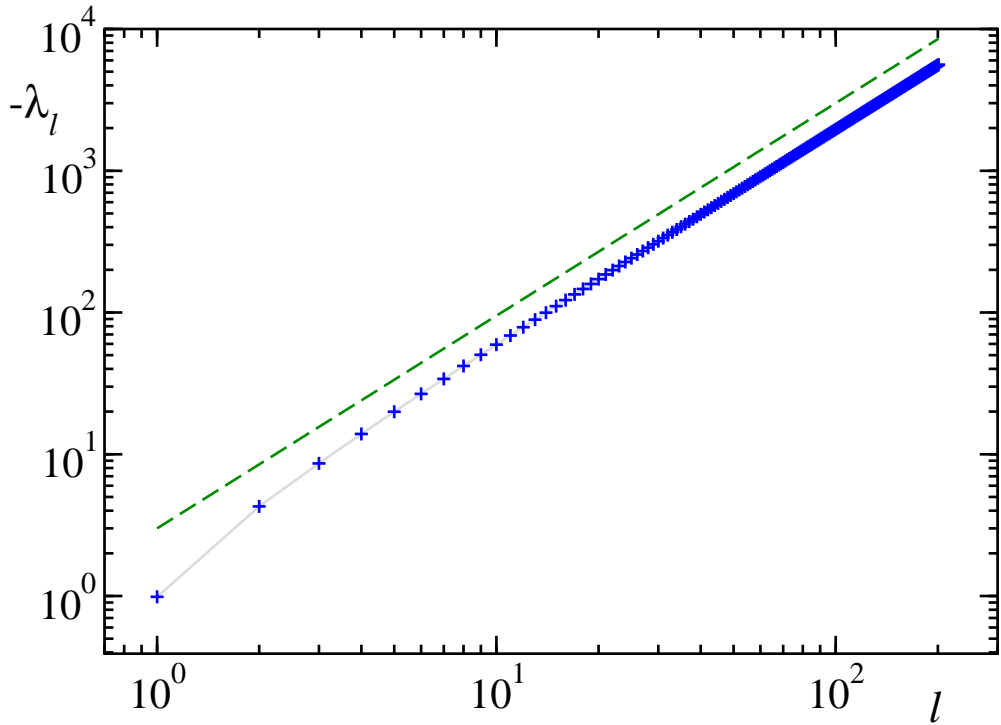}
\caption{Spectrum $\{\lambda_\ell\}$ of the linear equation \eref{eq:eraora}.
  The eigenvalues are expressed in $\e^3\w^2/\g$ time units. The straight
  line corresponds to a power law with a rate 3/2.
}
\label{fig1}
\end{center}
\end{figure}

A numerical evaluation reveals that ${\bf RDR}$ is almost diagonal. In
fact, the eigenvectors are very close to Fourier sine-modes
\cite{partII}.  In figure \ref{fig1}, we report the eigenvalues in
ascending order, versus the index $\ell$ that is equal (within a
proportionality factor) to the corresponding wave number.  The data
align almost perfectly along a straight line (in log-log scales) that
corresponds to a scaling with a power $3/2$.  The deviations observed
at small wave numbers are not due to the truncation of the operator
${\bf RDR}$; they express the fact that ${\bf RDR}$ is intrinsically
defined on a finite domain.  In \cite{partII}, we show that the
numerical solution of the entire dynamical operator (without any
approximation) confirms our analytical predictions.

Altogether, the space-time scaling of \eref{eq:eraora} indicates that the
evolution of the temperature field $T(y,t)$ is, on the considered time-scales,
ruled by a diffusion equation with a fractional Laplacian $\nabla^{3/2}$.
Recently, this has been shown to be the case for a similar model
system \cite{JKO08}, directly in the infinite-$N$ limit (i.e. without
including the effect of the boundary conditions).

%%%%%%%%%%%%%%%%%%%%%%%%%%%%%%%%%%%%%%%%%%%%%%%%%%%%%%%%%%%%%%%%%%%%%%%%%%%%%%%
\section{Discussion and conclusions}
\label{sec:concl}

We  have presented a detailed  description of the relaxation towards the
nonequilibrium steady state in a model of harmonic oscillators with
conservative noise. To our knowledge, this is an almost unique instance where
relaxation phenomena can be studied in great detail in a realistic setup.
By implementing the continuum-limit ideas previously
introduced in \cite{LMMP09} we have obtained a set of partial differential
equations describing the evolution of the covariance matrix. In the bulk, the
velocity-velocity and the symmetric component of the velocity-position
correlations are the relevant (slow) variables: they appear to evolve on
a time scale of order $1/\e^2 =N$. This means that in the bulk, relaxation
phenomena are mostly controlled by the propagation of sound waves.

Along the boundaries, the evolution of the relevant two-point
correlators can be explicitly determined to lowest order in $\e$.
Again, these correlators evolve on different time scales, the
temperature field being the slowest one (its dynamical equation
evolves on a time scales $t\sim N^{3/2}$).  By adiabatically
eliminating the fast variables, we find that $T(y,t)$ satisfies an
energy continuity equation, where the expression for the current can
be obtained from the stationary solution of the bulk equation.
Altogether, the temperature field $T(y,t)$ appears to satisfy a
diffusion equation with a fractional Laplacian $\nabla^{3/2}$.
However, the relationship with fractional Brownian motion should be
further explored. In particular, it is not yet clear to what extent
the temperature profile can be obtained from the solution of the
fractional equation in a finite domain.

The case of free BC remains open in view of the difficulties arising
along the boundaries, where the mathematical conditions depend
explicitly on system parameters such as $\omega$ and $\lambda$. This
is not only the indication of a lack of ``universality'' but implies
also that some variables (namely $\psi$) must scale differently in the
bulk and along the boundaries.  As a result, boundary layers are
expected to arise (and this is confirmed by the numerical analysis
carried out in the companion paper \cite{partII}) which would require
a separate analysis. This is one of the open problems that will be
worth investigating in the future, especially in the perspective that
a similar scenario might hold in generic nonlinear deterministic
systems. Only in the resonant case $\omega^2 = \lambda^2$, boundary
layers do not exist. However, even in this limit, the mathematical
conditions holding on the boundaries are sufficiently complicate to
prevent the derivation of explicit expressions (at least, to the best
of our knowledge).

All of our analysis has been {\it restricted} to two-point correlators. The main
reason is that the dynamical equations are exactly closed onto themselves, so
that there is no need to invoke higher-order correlators. Moreover, this
analysis allows determining exact expressions for the most relevant variables
such as the heat flux and the temperature profile. However, one should not
forget that the scaling behaviour of heat conductivity in this stochastic
model ($\kappa\simeq N^{1/2}$) differs from that of generic nonlinear systems,
where $\kappa\simeq N^{1/3}$. Is this an indication that a faithful description
of such systems needs including higher-order correlators? More modestly, it
would be already interesting to check to what extent a Gaussian approximation
of the invariant measure based on the knowledge of two-point correlators can
accurately describe other variables such as, e.g., energy fluctuations.

\appendix
%%%%%%%%%%%%%%%%%%%%%%%%%%%%%%%%%%%%%%%%%%%%%%%%%%%%%%%%%%%%%%%%%%%%%%%%%%%%%%%
\section{Derivation of the differential equations: bulk dynamics}
\label{ap:diff-bulk}

In  this  appendix  we  derive  the  set  of  partial  differential  equations
describing the dynamics  of the covariance matrix elements in  the bulk of the
system, namely for covariances $\mathbf{m}_{i,j}$ for which
\begin{enumerate}
\item the index of the momentum variable is in $[2,N-1]$,
\item the index of  $\Delta q$ is in $[3,N-1]$ for free  BC and in $[2,N]$ for
  fixed BC,
\item and $|i-j|>1$.
\end{enumerate}
In this situation the stochastic collision matrix is simply
\begin{equ} \label{Wbulk}
\mW_{ij} = \mV_{i+1,j} + \mV_{i,j-1} + \mV_{i-1,j} + \mV_{i,j+1} - 4\mV_{i,j} \ .
\end{equ}

In  the continuum  limit, discrete  shifts of  the indexes  $i$ and  $j$ yield
infinitesimal  changes  of  the field  variables  $x$  and  $y$ and  by  using
\eref{rule}, the set of difference equations \eref{eq:A}-\eref{eq:D} lead to a
set of continuous equation given explicitly by
\numparts
\begin{eqnarray}
\fl \dot\psi\left(x,y \right) = 
\omega^2\big[
-4 \mZ^+\left(x,y \right)
 + \mZ^+\left(x-\e,y-\e^2( y-1) \right)
 + \mZ^+\left(x+\e,y+\e^2( y+1) \right) \nonumber\\
\fl + \mZ^+\left(x-\e,y-\e^2( y+1) \right)
 + \mZ^+\left(x+\e,y+\e^2( y-1) \right)
 - \mZ^-\left(x-\e,y-\e^2( y-1) \right) \nonumber\\
\fl + \mZ^-\left(x+\e,y+\e^2( y+1) \right)
 - \mZ^-\left(x-\e,y-\e^2( y+1) \right)
 + \mZ^-\left(x+\e,y+\e^2( y-1) \right)
 \big] \nonumber\\
\fl + \g \left[\right.
 \mV\left(x+\e,y+\e^2( y+1) \right)
 + \mV\left(x-\e,y-\e^2( y-1) \right)
 + \mV\left(x-\e,y-\e^2( y+1) \right) \nonumber\\
\fl + \mV\left(x+\e,y+\e^2( y-1) \right)
 -4 \mV\left(x,y \right)
 \left.\right]  \ , \label{bulk-cont:A} \\
 \nonumber\\
\fl 2 \dot\mZ^-\left(x,y \right)   = 
 \g \big[
  -4\mZ^-\left(x,y \right)
 + \mZ^-\left(x-\e,y-\e^2( y-1) \right)
 + \mZ^-\left(x+\e,y+\e^2( y+1) \right) \nonumber\\
\fl + \mZ^-\left(x-\e,y-\e^2( y+1) \right)
 + \mZ^-\left(x+\e,y+\e^2( y-1) \right)
 + \mZ^+\left(x-\e,y-\e^2( y-1) \right) \nonumber\\
\fl - \mZ^+\left(x+\e,y+\e^2( y+1) \right)
 - \mZ^+\left(x-\e,y-\e^2( y+1) \right)
 + \mZ^+\left(x+\e,y+\e^2( y-1) \right)
\big]   \nonumber\\
\fl + \psi\left(x+\e,y+\e^2( y+1) \right)
 - \psi\left(x-\e,y-\e^2( y-1) \right)
 + \mV\left(x-\e,y-\e^2( y-1) \right) \nonumber\\
\fl - \mV\left(x+\e,y+\e^2( y+1) \right)
 - \mV\left(x-\e,y-\e^2( y+1) \right)
 + \mV\left(x+\e,y+\e^2( y-1) \right)\ , \label{bulk-cont:B} \\
 \nonumber\\
\fl  2 \dot\mZ^+\left(x,y \right)   = 
 \gamma  \big[
 -4 \mZ^+\left(x,y \right)
 + \mZ^+\left(x-\e,y-\e^2( y-1) \right)
 + \mZ^+\left(x+\e,y+\e^2( y+1) \right) \nonumber\\
\fl  + \mZ^+\left(x-\e,y-\e^2( y+1) \right)
 + \mZ^+\left(x+\e,y+\e^2( y-1) \right)
 + \mZ^-\left(x-\e,y-\e^2( y-1) \right) \nonumber\\
\fl  - \mZ^-\left(x+\e,y+\e^2( y+1) \right)
 - \mZ^-\left(x-\e,y-\e^2( y+1) \right)
 + \mZ^-\left(x+\e,y+\e^2( y-1) \right)\big]  \nonumber\\
\fl  + 2 \psi\left(x,y \right)
 - \psi\left(x+\e,y+\e^2( y+1) \right)
 - \psi\left(x-\e,y-\e^2( y-1) \right) \nonumber\\
\fl  + \mV\left(x-\e,y-\e^2( y-1) \right)
 +   \mV\left(x+\e,y+\e^2( y+1) \right) \nonumber\\
\fl  - \mV\left(x-\e,y-\e^2( y+1) \right)
 - \mV\left(x+\e,y+\e^2( y-1) \right) \ , \label{bulk-cont:C} \\
\nonumber\\
\fl  \dot\mV\left(x,y \right)  =  
 \omega^2\left[\right.
 - 2\mZ^+\left(x,y \right)
 + \mZ^+\left(x-\e,y-\e^2( y-1) \right)
 + \mZ^+\left(x+\e,y+\e^2( y+1) \right) \nonumber\\
\fl  - \mZ^-\left(x-\e,y-\e^2( y-1) \right) 
 + \mZ^-\left(x+\e,y+\e^2( y+1) \right)
 \left.\right] +
 \g \big[
   \mV\left(x+\e,y+\e^2( y+1) \right) \nonumber\\
\fl  + \mV\left(x-\e,y-\e^2( y-1) \right)
 + \mV\left(x-\e,y-\e^2( y+1) \right)
 + \mV\left(x+\e,y+\e^2( y-1) \right) \nonumber\\
\fl  -4 \mV\left(x,y \right)
 \big] \ . \label{bulk-cont:D}
\end{eqnarray}
\endnumparts  Finally,  straight  forward  differentiation in  the  continuous
coordinates  $x$  and  $y$, up  to  $\Or(\e^2)$,  lead  to  a set  of  partial
differential  equations for  the  time evolution  of  these four  correlators:
\numparts
\begin{eqnarray}
\dot{\psi } =  
 4\e\omega^2\mZ^-_x + 2\e^2\left(
 \omega^2\left(\mZ^+_{xx} +2y\mZ^-_{y}\right)
 +\gamma  \mV  _{xx}\right) \ ,  \label{bulk-diff:A}\\
\dot{\mZ^-} =
  \e\psi _x + \e^2\left(\gamma  \mZ^-_{xx} + y\psi _{y}
 \right) \ ,  \label{bulk-diff:B}\\ 
\dot{\mZ^+} =  
 \e^2\left(- \frac{1}{2}\psi _{xx}
 + \gamma\mZ^+_{xx} - \psi _{y} + 2\mV  _{y} \right) \ ,  \label{bulk-diff:C}\\
\dot{\mV} =  
 2\e\omega^2\mZ^-_x + \e^2\left(
 \omega^2\left(\mZ^+_{xx} + 2y\mZ^-_{y} + 2\mZ^+_{y}\right) + 2\gamma\mV_{xx}
 \right) \ . \label{bulk-diff:D} 
\end{eqnarray}
\endnumparts

%%%%%%%%%%%%%%%%%%%%%%%%%%%%%%%%%%%%%%%%%%%%%%%%%%%%%%%%%%%%%%%%%%%%%%%%%%%%%%%
\section{Derivation of the differential equations: $x=0$.}
\label{ap:diff-diag}

In this  appendix, we derive a  set of partial differential  equations for the
evolution of the diagonal covariances.  The dynamics in the boundary $\{x=0\}$
is different  from the bulk dynamics  due to the stochastic  collisions and is
not related to the {\it  physical} boundary conditions concerning the coupling
with the heat baths.

The dynamics  along this boundary  are obtained by considering  the difference
equations \eref{eq:A}-\eref{eq:D}  along the  diagonal ($i=j$), and  along the
the sub-diagonal ($i-j=1$).  Along the diagonal the difference equations become
\numparts
\begin{eqnarray}
\fl \dot\Omega_{i} =
2\w^2\big(\!-2\mS^+_{i,i}\!+\!\mZ^+_{i+1,i}\!+\!\mZ^+_{i,i-1}\!
 +\!\mZ^-_{i+1,i}\!+\!\mZ^-_{i,i-1}\big) + \gamma\big(T_{i-1}\!
 +\!T_{i+1}\!-\!2T_{i}\big) \ ,  \label{diag-d-A}\\ 
\fl \dot\mS^+_{i}\! =\!
\g\big(\!\!-\!2\mS^+_{i}\!+\!\mZ^+_{i+1,i}\!+\!\mZ^+_{i,i-1}\! 
 -\!\mZ^-_{i+1,i}\!+\!\mZ^-_{i,i-1}\!\big)\! +\! \big(\!\Omega_{i}\!
 -\!\psi_{i+1,i}\!+\!\mV_{i+1,i}\!-\!\mV_{i,i-1}\!\big) \ ,  \label{diag-d-C}
 \\ 
\fl \dot T_{i} =  2\w^2\big(\!-\mS^+_{i}\!+\!\mZ^+_{i+1,i}\!
 +\!\mZ^-_{i+1,i}\!\big)+\g\big(T_{i-1}\!+\!T_{i+1}\!
 -\!2T_{i}\!\big) \ . \label{diag-d-D}
\end{eqnarray}
\endnumparts

The continuum  limit rule \eref{rule}  for the diagonal on  sub-diagonal matrix
elements can be written as
\begin{equation} \label{rule-0}
\mathbf{m}_{i+\Delta  i,i+\Delta j} = \mathbf{m}(f\e,-1  + \e^2(-fy+s)) \ ,
\end{equation}
with the  shift functions $f$ and $s$  defined as before.  By  using this rule
and  differentiating  the resulting  continuous  equations,  we  obtain up  to
$\Or(\e^2)$, \numparts
\begin{eqnarray}
\fl \dot{\Omega} =  
 4\w^2\big(\mZ^+-\mS^++\mZ^-\big)+4\e\w^2\big(\mZ^+_x+\mZ^-_x\big)
 +2\e^2\w^2\big(\mZ^+_{xx}+\mZ^-_{xx}+2y(\mZ^+_{y}+\mZ^-_{y})\big) \ ,
 \label{diag-diff-A}\\
\fl \dot{\mS^+} =  
 \Omega-\psi + 2\g\big(\mZ^+-\mS^+\big) + \e\big(-\psi _x+2\g\mZ^+_x\big) +
 \e^2\big(-\frac{1}{2}\psi_{xx} + \g\mZ^+_{xx}\nonumber\\
 -(y+1)\psi_{y} + 2\g(y\mZ^+_{y}-\mZ^-_{y}) + 2\mV_{y}\big)
 \ , \label{diag-diff-C}\\ 
\fl \dot{T} =  
 2\w^2\big(\!\mZ^+\!-\!\mS^+\!+\!\mZ^-\!\big) + 2\e\w^2\big(\mZ^+_x+\mZ^-_x\big) 
 + \e^2\w^2\big(\mZ^+_{xx}+\mZ^-_{xx}+2(y+1)(\mZ^+_{y}+\mZ^-_{y})\big) \ . 
\label{diag-diff-D}
\end{eqnarray}
\endnumparts

Analogously, on the lower  diagonal ($i-j=1$), the difference equations become
\numparts
\begin{eqnarray}
\fl \dot\psi_{i,i-1} =
 \w^2\big(-4\mZ^+_{i,i-1} + \mS^+_{i} + \mZ^+_{i+1,i-1} + \mS^+_{i-1}
 + \mZ^+_{i,i-2} + \mZ^-_{i+1,i-1} + \mZ^-_{i,i-2} \big) \nonumber\\
 + \g\big(\mV_{i+1,i-1} + \mV_{i,i-2} - 2 \mV_{i,i-1} \big)
 \ , \label{ldiag-d-A} \\ 
\fl 2\dot\mZ^-_{i,i-1} =
 \g\big(-4\mZ^-_{i,i-1} + \mZ^-_{i+1,i-1} + \mZ^-_{i,i-2} + \mS^+_{i}
 - \mZ^+_{i+1,i-1} - \mS^+_{i-1,i-1} + \mZ^+_{i,i-2} \big) \nonumber\\
 + \psi_{i+1,i-1} - \Omega_{i} + T_{i}
 - \mV_{i+1,i-1} - T_{i-1} + \mV_{i,i-2} \ , \label{ldiag-d-B}\\
\fl 2\dot\mZ^+_{i,i-1} =
 \g\big(-4\mZ^+_{i,i-1} + \mS^+_{i} + \mZ^+_{i+1,i-1} + \mS^+_{i-1}
 + \mZ^+_{i,i-2} - \mZ^-_{i+1,i-1} + \mZ^-_{i,i-2} \big) \nonumber\\
 + 2\psi_{i,i-1} - \psi_{i+1,i-1} - \Omega_{i}
 + T_{i} + \mV_{i+1,i-1} - T_{i-1} - \mV_{i,i-2} \ , \label{ldiag-d-C}  \\
\fl \dot\mV_{i,i-1} =
 \w^2\big(-2\mZ^+_{i,i-1} + \mS^+_{i} + \mZ^+_{i+1,i-1} + \mZ^-_{i+1,i-1}\big)
 + \g\big(\mV_{i+1,i-1} + \mV_{i,i-2} - 2 \mV_{i,i-1}\big) \ , \label{ldiag-d-D} 
\end{eqnarray}
\endnumparts and by  using \eref{rule-0} and keeping differential  terms up to
$\Or(\e^2)$ we arrive at \numparts
\begin{eqnarray}
\fl \dot{\psi } =
 2\w^2\big(-\mZ^++\mS^++\mZ^-\big) + 2\e\big(2\omega^2\mZ^-_x
 +\g\mV_x\big)+ \e^2\big(2\w^2\big(\mZ^+_{xx}+2\mZ^-_{xx}+\mZ^+_{y}
 -\mS^+_{y} \nonumber\\
 +\big(2y-1\big)\mZ^-_{y}\big) +
 \g\big(3\mV_{xx}+2y\mV_{y}\big)\big) \ ,\label{ldiag-diff-A}\\ 
\fl \dot{\mZ^-} = 
 \frac{1}{2}\big(\psi\!-\!\Omega\big)-\gamma\mZ^- + \e\psi _x
 +\e^2\big(\psi_{xx}\!+\!y\psi_{y}\!-\!\mV_{y}\!+\!T_{y}
 +\g\big(\!\mZ^-_{xx}\!-\!\mZ^+_{y}\!+\!\mS^+_{y}\!+\!\mZ^-_{y}\!\big)
 \big) \ ,\label{ldiag-diff-B}\\
\fl \dot{\mZ^+} =
 \frac{1}{2}\big(\psi-\Omega\big)-\gamma\big(\mZ^+-\mS^+\big)
 +\e^2\big(\!\!-\!\frac{1}{2}\psi_{xx}\!-\!\psi_{y}\!+\!\mV_{y}\!+\!T_{y}
 +\g\big(\!\mZ^+_{xx}\!+\!\mZ^+_{y}\!-\!\mS^+_{y}\!-\!\mZ^-_{y}\!\big)
 \big) \ ,\label{ldiag-diff-C}\\
\fl \dot{\mV  } = 
 \omega^2\big(-\mZ^++\mS^++\mZ^-\big) + 2\e\big(\omega^2\mZ^-_x
 +\gamma\mV_x\big) + \e^2\big(\omega^2\big(\mZ^+_{xx}+2\mZ^-_{xx}
 +2\mZ^+_{y}+2y\mZ^-_{y}\big)\nonumber\\
 +\g\big(3\mV_{xx}+2y\mV_{y}\big)\big) \ .\label{ldiag-diff-D}
 \end{eqnarray}
\endnumparts

The set of partial differential equations
\eref{diag-diff-A}-\eref{diag-diff-D} and
\eref{ldiag-diff-A}-\eref{ldiag-diff-D}, describe the dynamics of the
covariance on the line $x=0$.

%%%%%%%%%%%%%%%%%%%%%%%%%%%%%%%%%%%%%%%%%%%%%%%%%%%%%%%%%%%%%%%%%%%%%%%%%%%%%%%
\section{Derivation of the differential equations: $y=\pm1$}
\label{ap:diff-border}

Along the boundaries, where the system is coupled with the heat baths, the
dynamics is different from that in the bulk not only because of the coupling
with the bath itself, but also because the deterministic force felt by the edge
oscillators feel from the bulk is not counter-balanced and, moreover, the
stochastic collision at the edges is a ``one-sided''process. Clearly, the
dynamics depend on whether the BC are free or fixed.

In this section  we derive the covariance dynamic  equations at the boundaries
$y=\pm1$. These boundaries  correspond, on the original matrix,  to the matrix
elements  of the first and last rows and  columns.  Restricted to the
semi-infinite  plane $x>0$,  {\it i.e.},  to the  domain  \eref{def:D}, $y=-1$
corresponds to  the first matrix column,  while $y=1$ corresponds  to the last
matrix row. Here  we specialize on the boundary $y=-1$. The  BC at $y=1$, that
can be  obtained analogously, lead to  the same information  that is extracted
from the BC at $y=-1$.

%%%%%%%%%%%%%%%%%%%%%%%%%%%%%%%%%%%%%%%%%%%%%%%%%%%%%%%%%%%%%%%%%%%%%%%%%%%%%%%
\subsection{Fixed boundary conditions: $y=-1$}
\label{ap:fixed}

We recall that fixed BC are defined by the relations $\Delta q_1 = q_1$ and
$\Delta  q_{N+1} =  -q_N$. Our starting point is transforming the set of
difference equations \eref{fixed-y-1:A1}-\eref{fixed-y-1:A3} into a set on the
variables  $\Psi$, $\mZ^\pm$  and $\mV$. Note however, that the domain of
$\Delta q_i$, which  is $i\in[1,N+1]$, is different from the domain of $p_i$,
$i\in[1,N]$. As a consequence, the terms $\mZ_{N+1,j}$ are non symmetrizable.

In what follows, we restrict to $i\in[3,N-1]$, where all terms are symmetrizable
and all covariances are ``far enough" from the diagonal. By finally denoting 
$\phi_{i,1}  \equiv \psi_{i,1}$, $\z^\pm_{i,1}  \equiv \mZ^\pm_{i,1}$
and $\nu_{i,1} \equiv \mV_{i,1}$, we obtain
\numparts
\begin{eqnarray}
 \fl \dot\phi_{i,1}  = 
 \w^2\big[
 -4 \z^+_{i,1}  + \z^+_{i+1,1} + \z^+_{i-1,1}
  + \z^-_{i+1,1} - \z^-_{i-1,1}+
  \mZ^+_{i,2}-\mZ^-_{i,2}\big]-\l\nu_{i,1}\nonumber\\ 
  + \g\big[\nu_{i+1,1} + \nu_{i-1,1} + \mV_{i,2} 
    -3\nu_{i,1} \big]   \ , \label{fbby-1:A} \\ 
\fl  2\dot\z^-_{i,1}  =  
 \g\big[
 -3\z^-_{i,1} + \z^-_{i+1,1} + \z^-_{i-1,1} 
 +\z^+_{i,1}- \z^+_{i+1,1} - \z^+_{i-1,1} + \mZ^-_{i,2} + \mZ^+_{i,2}
 \big] \nonumber \\ -\l\big[\z^+_{i,1}+\z^-_{i,1}\big]
  + \phi_{i+1,1} - \psi_{i,2}  - \nu_{i+1,1} - \nu_{i-1,1}+ \mV_{i,2}
 \ , \label{fbby-1:B}\\
\fl  2\dot\z^+_{i,1}  =  
 \g\big[
 -3\z^+_{i,1} + \z^+_{i+1,1} + \z^+_{i-1,1} 
 +\z^-_{i,1}- \z^-_{i+1,1} - \z^-_{i-1,1} + \mZ^+_{i,2} + \mZ^-_{i,2}
 \big] \nonumber \\ -\l\big[\z^+_{i,1}+\z^-_{i,1}\big]
  +2\phi_{i,1}+ \phi_{i+1,1} - \psi_{i,2}  + \nu_{i+1,1} - \nu_{i-1,1}+ \mV_{i,2}
 \ , \label{fbby-1:C}\\
 \fl  \dot\nu_{i,1}   =  \w^2\big[
 -2\z^+_{i,1} + \z^+_{i+1,1} + \z^-_{i+1,1} + \mZ^+_{i,2} -
 \mZ^-_{i,2}\big] -\l\nu_{i,1}\nonumber  \\ 
  + \g\big[\nu_{i+1,1} + \nu_{i-1,1} + \mV_{i,2} -3\mV_{i,2}\big]
  \ . \label{fbby-1:D}   \\
 \fl \dot\psi_{i,2}  = 
 \w^2\big[
 -4 \mZ^+_{i,2} + \mZ^+_{i,3} + \mZ^+_{i+1,2} + \mZ^+_{i-1,2} + \z^+_{i,1}
 - \mZ^-_{i,3} + \mZ^-_{i+1,2} - \mZ^-_{i-1,2} + \z^-_{i,1} \big]\nonumber\\
  + \g\big[\mV_{i+1,2} + \mV_{i,3} + \mV_{i-1,2} + \nu_{i,1}
    -4\mV_{i,2} \big]   \ , \label{fbby-1:E} \\  
\fl  2\dot\mZ^-_{i,2}  =  
 \g\big[
 -4 \mZ^-_{i,2} + \mZ^-_{i,3} + \mZ^-_{i+1,2} + \mZ^-_{i-1,2} + \z^-_{i,1}
 + \mZ^+_{i,3} - \mZ^+_{i+1,2} - \mZ^+_{i-1,2} + \z^+_{i,1} \big]\nonumber \\
  + \psi_{i+1,2} - \psi_{i,3} + \mV_{i,3} - \mV_{i+1,2} - \mV_{i-1,2}
 + \nu_{i,1}   \ , \label{fbby-1:F}\\
\fl 2\dot\mZ^+_{i,2}  =  \g\big[
 -4 \mZ^+_{i,2} + \mZ^+_{i,3} + \mZ^+_{i+1,2} + \mZ^+_{i-1,2} + \z^+_{i,1}
 + \mZ^-_{i,3} - \mZ^-_{i+1,2} - \mZ^-_{i-1,2} + \z^-_{i,1}
 \big] \nonumber\\
  + 2 \psi_{i,2} - \psi_{i+1,2} - \psi_{i,3} + \mV_{i,3} + \mV_{i+1,2}
 - \mV_{i-1,2} - \nu_{i,1}  \ , \label{fbby-1:G}\\
 \fl  \dot\mV_{i,2}   =  \w^2\big[
 -2 \mZ^+_{i,2} + \mZ^+_{i,3} + \mZ^+_{i+1,2} - \mZ^-_{i,3} + \mZ^-_{i+1,2}
 \big]\nonumber  \\
  + \g\big[\mV_{i+1,2} + \mV_{i,3} + \mV_{i-1,2} + \nu_{i,1}
    -4\mV_{i,2} \big]  \ . \label{fbby-1:H}  
\end{eqnarray}
\endnumparts
In the continuum limit, the column index $1$ corresponds to $y=-1$. As a result,
\eref{rule} becomes
\begin{equation} \label{rule-1}
\mathbf{m}_{i+\Delta  i,1+\Delta j} = \mathbf{m}(x+f\e,-1  + \e^2(-f+s)) \ ,
\end{equation}
with  the shift  functions $f$  and  $s$ defined  as before.   Using this  and
proceeding as  in \ref{ap:diff-bulk}, the dynamics at $y=-1$ for fixed BC is
described by the following set of partial differential equations \numparts
\begin{eqnarray}
\fl \dot\phi = \w^2\big(\mZ^+\!-\!\mZ^-\! -\! 2\z^+ \big) -\l\nu -\g\big(\nu\! -
\!\mV\big) + \e\big[-\w^2\mZ^+_x - \g\mV_x\big] \ , \label{fbby-3:A}\\ 
\fl 2\dot{\z}^- = \g\big(\mZ^+\!+ \!\mZ^-\! -\! \z^+\! -\! \z^- \big)
-\l\big(\z^+\!+\!\z^-\big) - \psi + \phi -2\nu + \mV - \e\big[\g\mZ^+_x \!-
  \!\mV_x\big]\ , \label{fbby-3:B}\\  
\fl 2\dot{\z}^+ = \g\big(\mZ^+\! +\! \mZ^-\! -\! \z^+\! -\! \z^-\big)
-\l\big(\z^+\!+\!\z^-\big) - \psi + \phi + \mV + \e\big[\!-\!\g\mZ^+_x + 2\nu_x
  \!-\!\mV_x\big] \ , \label{fbby-3:C}\\
\fl \dot{\nu} = \w^2\big(\mZ^+\!-\! \mZ^-\! -\! \z^+ \!+\! \z^- \big)
  -\l\nu -\g\big(\nu\! -\!\mV\big) +\e\big[\w^2\big(\z^+_x\! - \mZ^+_x\big)\! -
    \!\g\mV_x\big] \ , \label{fbby-3:D}\\ 
\fl \dot{\psi} = \w^2\big(\!-\!\mZ^+\!-\!\mZ^-\!+\!\z^+\!+\!\z^-\big) +
  \g\big(\nu\!-\!\mV) +4\e\w^2\mZ^-_x +
  2\e^2\big[\w^2\mZ^+_{xx} + \g \mV_{xx}\big]  \ ,  \label{fbby-3:E}\\  
\fl 2\dot{\mZ}^- = \g\big(\!-\!\mZ^+\!-\!\mZ^-\!+\!\z^+\!+\!\z^-\big) +
  \nu\!-\!\mV +2\e\psi_x \ , \label{fbby-3:F}\\   
\fl 2\dot{\mZ}^+ = \g\big(\!-\!\mZ^+\!-\!\mZ^-\!+\!\z^+\!+\!\z^-\big) -
  \nu\!+\!\mV + 2\e^2\big[\g\mZ^+_{xx}+2\mV_y\big]
  \ , \label{fbby-3:G}\\  
\fl \dot{\mV} = \g\big(\nu-\mV) +2\e\w^2\mZ^-_x +
\e^2\big[\w^2\big(\mZ^+_{xx}+ 2\mZ^+_y \big) +
  2\g\mV_{xx}\big]  \ , \label{fbby-3:H}    
\end{eqnarray}
\endnumparts

%%%%%%%%%%%%%%%%%%%%%%%%%%%%%%%%%%%%%%%%%%%%%%%%%%%%%%%%%%%%%%%%%%%%%%%%%%%%%%%
\subsection{Free boundary conditions: $y=-1$}
\label{ap:free}

We start by  recalling that free BC are defined  by setting $\Delta q_1=\Delta
q_{N+1}=0$. This means that in this case, the domain of the phase variables is
$i\in[2,N]$ for  $\Delta q_i$ and $i\in[1,N]$ for  $p_i$.  Consequently, while
$\mZ_{i>1,1}$ is well defined, its symmetric component $\mZ_{1,i}$ is not, thus
restricting  the  symmetrization  of  the  covariance  $\mZ$  \eref{Zsym}.  At
variance with the  case of  fixed  BC, here  it is  necessary to  consider
non-symmetrizable terms.  In analogy to the  previous section, we  distinguish
boundary covariances from their bulk counterparts by defining
$\nu_{i,1}\equiv\mV_{i,1}$  and  $\z_{i,1}\equiv\mZ_{i,1}$,  recalling that
$\z_{i,1}$ is  non-symmetrizable. Moreover, note that for free BC, $\psi$
has no boundary component, as the dynamics of $\psi_{i,2}$ corresponds to that
in the bulk (see \eref{free-y-1:A1}).

By transforming the set of difference equations
\eref{free-y-1:A1}-\eref{free-y-1:A3} into the covariance variables of section
\ref{sec:change} and restricting to $i\in[4,N-1]$, we obtain \numparts
\begin{eqnarray}
 \fl \dot\psi_{i,2}  = 
 \omega^2\big[
 -4 \mZ^+_{i,2} + \mZ^+_{i,3} + \mZ^+_{i+1,2} + \mZ^+_{i-1,2}
 - \mZ^-_{i,3} + \mZ^-_{i+1,2} - \mZ^-_{i-1,2} + \z_{i,1} \big]\nonumber\\
  + \gamma  \left[\mV_{i+1,2} + \mV_{i,3} + \mV_{i-1,2} + \nu_{i,1}
    -4\mV_{i,2} \right]   \ , \label{bby-1:A} \\  
\fl  2\dot\mZ^-_{i,2}  =  
\gamma  \left[
 -4 \mZ^-_{i,2} + \mZ^-_{i,3} + \mZ^-_{i+1,2} + \mZ^-_{i-1,2}
 + \mZ^+_{i,3} - \mZ^+_{i+1,2} - \mZ^+_{i-1,2} + \z_{i,1} \right]\nonumber \\
  + \psi_{i+1,2} - \psi_{i,3} + \mV_{i,3} - \mV_{i+1,2} - \mV_{i-1,2}
 + \nu_{i,1}   \ , \label{bby-1:B}\\
\fl 2\dot\mZ^+_{i,2}  =  \gamma  \left[
 -4 \mZ^+_{i,2} + \mZ^+_{i,3} + \mZ^+_{i+1,2} + \mZ^+_{i-1,2}
 + \mZ^-_{i,3} - \mZ^-_{i+1,2} - \mZ^-_{i-1,2} + \z_{i,1}
 \right] \nonumber\\
  + 2 \psi_{i,2} - \psi_{i+1,2} - \psi_{i,3} + \mV_{i,3} + \mV_{i+1,2}
 - \mV_{i-1,2} - \nu_{i,1}  \ , \label{bby-1:C}\\
\fl  \dot\mV_{i,2}   =  \omega^2\left[
 -2 \mZ^+_{i,2} + \mZ^+_{i,3} + \mZ^+_{i+1,2} - \mZ^-_{i,3} + \mZ^-_{i+1,2}
 \right]\nonumber  \\
  + \gamma  \left[\mV_{i+1,2} + \mV_{i,3} + \mV_{i-1,2} + \nu_{i,1}
    -4\mV_{i,2} \right]  \ . \label{bby-1:D}  \\
\fl \dot\z_{i,1}  =  
\gamma  \left[\mZ^+_{i,2} + \mZ^-_{i,2} - \z_{i,1}\right] -\l\z_{i,1}
  - \psi_{i,2} + \nu_{i,1} - \nu_{i-1,1} + \mV_{i,2}  \ , \label{bby-1:E}\\
\fl  \dot\nu_{i,1}   =  \omega^2\left[\z_{i+1,1} - \z_{i,1} + \mZ^+_{i,2} -
  \mZ^-_{i,2}\right] -\l\nu_{i,1} + \g\left[\nu_{i+1,1} + \nu_{i-1,1} +
  \mV_{i,2} + - 3\nu_{i,1} \right]  \ . \label{bby-1:F}
\end{eqnarray}
\endnumparts

Furthermore,  using \eref{rule-1},  we obtain  the continuous  version  of the
equations above. Differentiating these later equations and keeping terms up to
$\Or(\e^2)$, we  obtain the following  set of partial  differential equations:
\numparts
\begin{eqnarray}
\fl \dot\psi = \w^2\big(\!-\mZ^-\!+\!\z\!-\!\mZ^+\!\big) +
\g\big(\nu\!-\!\mV\big) + 4\e\w^2\mZ^-_x +
2\e^2\big[\w^2\mZ^+_{xx}+\g\mV_{xx}\big]
\ ,\label{bby-3:A}\\    
\fl 2\dot{\mZ}^- = \g\big(-\!\mZ^-\!+\!\z\!-\!\mZ^+\!\big) + \nu\!-\!\mV 
+2\e\psi_x 
\ ,\label{bby-3:B}\\ 
\fl 2\dot{\mZ}^+ = \g\big(-\!\mZ^-\!+\!\z\!-\!\mZ^+\!\big) - \nu+\mV 
+ 2\e^2(\w^2\mZ^+_{xx}+2\mV_y)   ,\label{bby-3:C}\\
\fl \dot{\mV} = \g\big(\nu\!-\!\mV\big) + 2\e\w^2\mZ^-_x
+ \e^2\big[\w^2(\mZ^+_{xx}  + 2\mZ^+_{y}) +
  2\g\mV_{xx}\big]\ ,\label{bby-3:D}\\ 
\fl \dot{\z} = -\g\big(\!-\!\mZ^-\!+\!\z\!-\!\mZ^+\!\big)-\l\z+\mV-\psi 
+ \e\big[-\g(\mZ^+_x+\mZ^-_x)+\nu_x-\mV_x+\psi_x\big] \ ,\label{bby-3:E}\\
\fl \dot{\nu} = \w^2\big(\mZ^+\!-\!\mZ^-\!\big) -\g\big(\nu-\mV\big) -\l\nu
+\e\big[\w^2(\mZ^-_x\!+\!\z_x\!-\!\mZ^+_x\!)-\g\mV_x\big] \ .\label{bby-3:F}
\end{eqnarray}
\endnumparts

We remark  that the covariance's equations  at $y=-1$ for  the diagonal terms,
namely those obtained for $i=1$ and  $i=2$, yield the same solution to leading
order, with the additional constraint $\nu_{i,1} = T_+$.

\ack
We acknowledge useful discussions with L. Delfini and R. Livi. This work is
partially supported by the the Italian project {\it Dinamiche cooperative in
strutture quasi uni-dimensionali} No. 827 within the CNR program Ricerca
spontanea a tema libero. C.M.-M. was partially funded by the European Research
Council and the Academy of Finland.

\section*{References}
\bibliographystyle{unsrt}

%\bibliography{paper-13}

\end{document}